\documentclass{aa}

\usepackage{natbib}
\usepackage{psfig}
\usepackage{txfonts}
\usepackage{graphicx}

\begin{document}

\title{Be/X-ray stars and candidates: catalogue}

\titlerunning{Be/X-ray stars}

\author{S.B. Popov
            \inst{1,2}
             \and
N.V. Raguzova\inst{1}
         }

   \offprints{N. Raguzova}

   \institute{Sternberg Astronomical Institute,
Universitetski pr. 13, 119992 Moscow, Russia\\
\email{polar@sai.msu.ru; raguzova@sai.msu.ru}
             \and
Universit\`a di Padova, Dipartimento di Fisica, 
via Marzolo 8, 35131, Padova, Italy\\
\email{popov@pd.infn.it}
}

   \date{}

\abstract{Here we present a compilation
of  data on Be/X-ray binaries. On the whole we include 90 objects
into our catalogue. Brief comments on each object are provided.
}

\maketitle

\section{Introduction}

Among massive X-ray binaries (HMXBs) most numerous are systems where
optical companions are Be-stars. The latest catalogue of HMXBs was
presented by \cite{liu2000}. However, as far as Be/X-ray systems are a
subject of particular interest it is reasonable to make a separate
list of these sources. Such lists appear more or less regularly
(see one of the latest one in Ziolkowski 2002\nocite{z2002}). In this paper
we try to collect a more extended catalogue and provide some brief
comments
on every source \footnote{This catalogue is appearing only in the ArXiv
and should be refered only by its astro-ph number. Comments are welcomed.}.

\section{The catalogue}
\label{catalogue}

In the tables below we present a compilative catalogue of Be/X-ray
stars.
In the first column we give sources names. If possible the first name
corresponds to notation in \cite{liu2000}.
In the second and third columns
we present spectral type of the massive companion and its magnitude.
In the forth column we give spin period, and in the fifth --- orbital
period. Then we give orbital eccentricity and distance to the source.
In the 8th column we give $L_\mathrm{max}$ -- the maximal observed luminosity.
In the last column  pulse fractions are given.
Some comments and more detailed description about each object
can be found in subsection "Comments to the tables" below.
References in the tables are given in square brackets
(in few cases we do not follow the data up to 
the first determination of a parameter, but give a 
reference to some catalogue).

\begin{table*}[t]
\caption{Be/X-Ray Binaries \label{table}
}
\begin{tabular}{|l|c|c|c|c|c|c|c|c|c|}
\hline

Name        & Spec. & m & $P_\mathrm{spin}$, s & $P_\mathrm{orb}$, d &  e
 & d, kpc & $L_\mathrm{max}$ & Pulse   \\
 & type    &      &                            &
 &       &              &  erg s$^{-1}$ & frac., \% \\

\hline

 & & & & & & & &  \\

XTE SMC 95 & & & 95 [1] & & & SMC & $2\cdot 10^{37}$ [1] & \\
J0032.9-7348 & Be & & & & & SMC & $1.3\cdot 10^{37}$ [2] & \\
J0049-732       & Be? & & 9.1320 [3] & & & SMC &  $4\cdot 10^{35}$ [3] &
\\
J0049-729       & Be & 16.92 [2] & 74.67 [4] & & & SMC & $7.5\cdot 10^{36}$
[5] & 70 [5] \\
J0049.4-7323  & Be & & 755.5 [6] & & & SMC & $5\cdot 10^{35}$ [7] & \\
0050-727        & O9 V-IIIe [8] & 14 [8] & 92 [9] & & & SMC & $6\cdot
10^{37}$ [10] &  \\
(SMC X-3)      & & & & & & & & \\
J0050.7-7316  & B0-B0.5 Ve [11] & 15.4 [11] & 323.1 [12] & & & SMC &
$1.8\cdot 10^{36}$ [2] & \\
(DZ Tuc)        & & & & & & & & \\
J0051-722        & Be & 15 [13] & 91.1 [2] & & & SMC & $2.9\cdot
10^{37}$
[2] &  \\
0051.1-7304  & Be & 14.28 [2] & & & & SMC & $1.6\cdot 10^{35}$ [2] &
\\
J0051.8-7231  & Be & 13.4 [2] & 8.9 [2] & & & SMC & $1.4\cdot 10^{36}$
[2] & 25 [14] \\
J0051.9-7311  & Be & 14.4 [2] & 172.4 [15] & & & SMC & $4.7\cdot
10^{35}$
[2] & \\
J0052-723        & B0V-B1Ve [16] & 15.8 [16] & 4.78 [17] & & & SMC &
$\sim 10^{38}$ [16] &  \\
J0052.1-7319  & O9.5 IIIe [18] & 14.6 [18] & 15.3 [2] & & & SMC &
$1.3\cdot 10^{37}$ [2] & \\
J0052.9-7158  & Be & 15.46 [19] & 167.8 [20] & & & SMC & $2.0\cdot
10^{37}$ [2] & \\
J0053.8-7226  & Be & & 46.6 [2] & & & SMC & $7.4\cdot 10^{36}$ [2] & 25
[21] \\
0053-739        & B1.5 Ve [22] & 16.0 [22] & 2.37 [23] & & & SMC &
$8.4\cdot 10^{37}$ [2] & \\
(SMC X-2)      & & & & & & & & \\
0053+604       & B0.5 IVe [24] & 1.6-3.0 [25] & & 203.59 [26] & 0.26 [26]
&
0.188 [27] & $3.9\cdot 10^{34}$ [2] & \\
($\gamma$ Cas) & & & & & & & & \\
J0054.9-7226   & B0-B1 III-Ve [8] & 15.28 [2] & 59.07 [2] & 65 [28] & &
SMC & $3.0\cdot 10^{37}$ [2] \\
J0057.4-7325   & Be? & & 101.45 [29] & & & SMC & $1.2\cdot 10^{36}$ [30]
& \\
J0058.2-7231   & B2-3 Ve [31] & 14.9 [2] & & & & SMC & $2.1\cdot
10^{35}$
[2] & \\
J0058-720         & Be? & & 281.1 [32] & & &  SMC & $1.6\cdot 10^{36}$
[2] & \\
J0059.2-7138   & B1 IIIe [8] & 14.08 [2] & 2.763 [2] & & & SMC &
$5.0\cdot 10^{37}$ [2] & 37 [33] \\
J0101.0-7206   & Be & & 304.49 [34] & & & SMC & $1.3\cdot 10^{36}$
[2] &       \\
J0101.3-7211   & Be & & 455 [35] & & & SMC & $7.3\cdot 10^{35}$ [36] &
\\
J0103-722         & O9-B1 III-Ve [8] & 14.80 [2] & 345.2 [2] & & & SMC &
$1.5\cdot 10^{36}$ [2] & \\
J0106.2-7205   & B2-5 III-Ve [8] & 16.7 [8] & & & & SMC & $5\cdot
10^{34}$ [37] & \\
0103-762         & Be & 17 [8] & & & & SMC & $4.3\cdot 10^{35}$ [38] &
\\
J0105-722         & Be? & & 3.343 [2] & & & SMC & $1.5\cdot 10^{35}$ [2]
& \\
J0111.2-7317   & B0.5-B1Ve [18] & 15.4 [18] & 31.03 [2] & & & SMC &
$2.0\cdot 10^{38}$ [2] & 45 [18] \\
0115+634        & B0.2 Ve [39] & 15.5 [39] & 3.6 [2] & 24.3 [39] &
0.34
[39] & 7-8 [39] & $3.0\cdot 10^{37}$ [2] & 40-60 [70]\\
J0117.6-7330   & B0.5 IIIe [40] & 14.2 [2] & 22.07 [2] & & & SMC &
$1.2\cdot 10^{38}$ [2] & 11.3 [41] \\
J0146.9+6121  & B1 III-Ve [42] & 11.2 [42] & 1412 [2] & & & 2.5 [2] &
$3.5\cdot 10^{35}$ [2] & \\
(V831 Cas)     & & & & & & & & \\
0236+610        & B0.5 Ve [43] & 10.7 [8] & & 26.45 [8] & & 3.1 [2] &
$2\cdot 10^{34}$ [2] & \\
(V615 Cas)     & & & & & & & & \\
0331+530        & O8-9 Ve [44] & 15.7 [44] & 4.4 [2] & 34.3 [44] & 0.3
[44] & 7 [44] & $ \ga 10^{38}$ [44] & \\
(BQ Cam)       & & & & & & & & \\
0352+309        & O9.5 IIIe & 6.1-6.8 [45] & 837 [45] & 250
[45] & 0.11[45] & 1.3 [45] & $3\cdot 10^{35}$ [45] &  \\
 (X Per)            & -B0 Ve [45]& & & & & & & \\
J0440.9+4431  & B0 III-Ve [36] & 10.78 [36] & 203 [2] & & & 3.2 [2] &
$3\cdot 10^{34}$ [2] & \\
J0501.6-7034   & B0 Ve [46] & 14.5 [46] & & & & LMC & $7\cdot 10^{34}$
[46] & \\
J0502.9-6626    & B0 Ve [46] & 14.3 [46] & 4.1 [46] & & & LMC & $4\cdot
10^{37}$ [46] & \\
J0516.0-6916    & B1 V [46] & 15.0 [8] & & & & LMC &$5\cdot 10^{35}$
[8] & \\
 & Be [76] & & & & & & & \\
J0520.5-6932    & O9 Ve [46] & 14.4 [8] & & 24.4 [77]
& & LMC & $8\cdot 10^{38}$ [77] &  \\
J0529.8-6556    & B0.5 Ve [4] & & 69.5 [47] & & & LMC & $2\cdot
10^{36}$ [47] & \\
053109-6609.2  & B0.7 Ve [46] & & 13.7 [46] & 24.5 [46] & & LMC &
$1\cdot 10^{37}$ [47] & 54-78 [47] \\
J0531.5-6518    & B2 Ve [46] & 16.02 [46] & & & & LMC & $3\cdot 10^{35}$
[46] & \\
J0535.0-6700    & B0 Ve [46] & 14.87 [46] & & & & LMC & $3\cdot 10^{35}$
[46] & \\
0535-668          & B0.5 IIIe [46] & 12.3-14.9 [8] & 0.068 [8] & 16.7
[8] & $> 0.5$ [24] & LMC & $1\cdot 10^{39}$ [46] & \\

\hline
\end{tabular}
\end{table*}

References for the table:\\
(1) \cite{lcp2002};
(2) \cite{hs00};
(3) \cite{uyi00a};
(4) \cite{yk98a};
(5) \cite{yit99};
(6) \cite{uyi00b};
(7) \cite{yiu00};
(8) \cite{liu2000};
(9) \cite{mlt97};
(10) \cite{ljc77};
(11) \cite{chl02};
(12) \cite{iyt99};
(13) \cite{scb99};
(14) \cite{isa95};
(15) \cite{yti00};
(16) \cite{lcc03};
(17) \cite{cmm01};
(18) \cite{cnc01};
(19) \cite{csm97};
(20) \cite{yit03};
(21) \cite{l98};
(22) \cite{mmt79};
(23) \cite{cm00};
(24) \cite{z2002};
(25) \cite{j03};
(26) \cite{hhs00};
(27) \cite{plk97};
(28) \cite{lwc99};
(29) \cite{ytk00};
(30) \cite{tky00};
(31) \cite{ec03};
(32) \cite{sph03};
(33) \cite{kyk00};
(34) \cite{mfl03};
(35) \cite{shk01};
(36) \cite{yit00};
(37) \cite{hs94};
(38) \cite{yit03};
(39) \cite{no01};
(40) \cite{s99};
(41) \cite{mfh98};
(42) \cite{rnc00};
(43) \cite{hc81};
(44) \cite{nrf99};
(45) \cite{dlp01};
(46) \cite{nc02};
(47) \cite{bsr98};
(48) \cite{nrf00};
(49) \cite{bks98};
(50) \cite{ffm85}.
(51) \cite{isc2001};
(52) \cite{pam2001};
(53) \cite{tcm1998};
(54) \cite{kth1998};
(55) \cite{whs1997};
(56) \cite{pg1994};
(57) \cite{rr99};
(58) \cite{wfs1999a};
(59) \cite{kph1999};
(60) \cite{cmn2000};
(61) \cite{wfs1999b};
(62) \cite{cp1997};
(63) \cite{icm2001};
(64) \cite{wfc2003};
(65) \cite{inc2001};
(66) \cite{whs1997};
(67) \cite{pss2000};
(68) \cite{rnb2001};
(69) \cite{rr1999a};
(70) \cite{hwf04}
(71) \cite{oop1999};
(72) \cite{kra1983};
(73) \cite{torii1999};
(74) \cite{smith1998};
(75) \cite{mhd1997};
(76) \cite{scc99};
(77) \cite{ecg2004};
(78) \cite{torii1998};
(79) \cite{rnf2004};
(80) \cite{icp2000}.

\begin{table*}[t]
\caption{Be/X-Ray Binaries (continue) \label{table}
}
\begin{tabular}{|l|c|c|c|c|c|c|c|c|}
\hline

Name        & Spectral & m & $P_\mathrm{spin}$, s & $P_\mathrm{orb}$, d &
  e  & d, kpc &  $L_\mathrm{max}$ & Pulse   \\
 & type& & & &  &  & & frac., \%   \\            

\hline

0535+262         & B0 IIIe [48] & 8.9-9.6 [8] & 105 [2] & 111 [8] &
0.47 [24] & 2.4 [2] & $2\cdot 10^{37}$ [2] & 20-100 \\
(V725 Tau)      & & & & & & & &  [49], [50] \\

0544-665          & B0 Ve [46] & 15.55 [46] & & & & LMC &$1\cdot
10^{37}$ [46] & \\

0544.1-710       & B0 Ve [46] & 15.25 [46] & 96.08 [46] & & & LMC
&$2\cdot 10^{36}$ [46] & \\
0556+286 & B5ne [8] & 9.2 [8]& & & & & &  \\

J0635+0533 & B2V-B1IIIe [8] & 12.83 [8]& 0.0338 [8]& & &  $2.5-5$
[59]
& $(9$-$35)\cdot 10^{33}$ [59] & $\sim$20 [60] \\

0726-260 & O8-9Ve [8] & 11.6 [8] & 103.2 [8] & 34.5 [8] & & $\sim$6 [62] &  
$2.8 \cdot 10^{35}$ [62] &  $\sim 30$ [62] \\

0739-529 & B7 IV-Ve [8]& 7.62 [8]& & & & & &   \\

0749-600 & B8 IIIe [8]& 6.73 [8]& & & & & &   \\

J0812.4-3114 &  B0.5 V-IIIe [8]& 12.42 [8]& 31.89 [8]& 80 [68] & & 9 [69] & 
$1.1\cdot 10^{36}$ [69] &  \\

0834-430 & B0-2 III-Ve [8]& 20.4 [8]& 12.3 [8]& 105.8 [24] & 0.12 [24] & 5 [2] &
 $1.1\cdot 10^{37}$[2] &  $<$15 [55] \\

J1008-57 & O9e-B1e [24] & 15.27 [8]& 93.5 [8]& 247.5 [24] & 0.66 [24] & 2 [2] &

$2.9 \cdot 10^{35}$ [2] &  60 [56] \\

1036-565 & B4 IIIe [8]& 6.64 [8]& & & & & &   \\

J1037.5-5647 & B0V-IIIe [8]& 11.3 [8]& 862 [8]& & & 5 [2] &
 $4.5\cdot 10^{35}$ [2]&  52 [57] \\

1118-615 & O9.5 V-IIIe [8]& 12.1 [8]& 405 [8]& & & 5 [2] &  $5\cdot 10^{36}$ [2]&
\\

1145-619 & B0.2 IIIe [24] & 9.3 [8]& 292.4 [8]& 187.5 [8]& $>$0.5 [24] & 0.5 [2] & $7.4 \cdot 10^{34}$ [2] &  28-70 [58] \\


1249-637 & B0 IIIe [8]& 5.31 [8]& & & & & &   \\

1253-761 & B7 Vne [8]& 6.49 [8]& & & & & &   \\

1255-567 & B5 Ve [8]& 5.17[8] & & & & & &   \\

1258-613 & B2 Vne [8]& 13.5 [8]& 272 [8]& 132.5 [24] & $>$0.5 [24] & 2.4 [2] &  
$1\cdot 10^{36}$ [2] &  \\
(GX 304-1) & & & & & & &  &  \\


1417-624 &  B1 Ve [24] & 17.2 [8]& 17.6 [8]& 42.12 [8]& 0.446 [24] & 10 [2] & 
 $8 \cdot 10^{36}$ [2]  &  \\

J1452.8-5949 &  & & 437.4 [8]& & & 9 [2] &  $8.7\cdot 10^{33}$ [2] &  
50-100 [71]\\

J1543-568 & B0.7 Ve [24] & & 27.1 [24] & 75.6 [63] & $<$0.03 [63] 
& $>10$ [63] & $>10^{37}$ [63] &   60-70 [63] \\

1553-542 & Be? & & 9.26 [24] & 30.6 [8]& $<$0.09 [24] & 10 [2] & 
$7 \cdot 10^{36}$ [2] & 30 [72] \\

1555-552 & B2nne [8]& 8.6 [8]& & & & & &  \\

J170006-4157 &  & & 714.5 [8]& & & 10 [2] &  $7.2 \cdot 10^{34}$ [2]&
$\sim 30$ [73]\\

J1739-302 &  & & & & & 8.5 [74] & $4.2\cdot 10^{37}$ [74]&   \\

J1739.4-2942 & Be? & & & & &  & &  \\

J1744.7-2713 & B2 V-IIIe [8] & 8.4 [8]& & & & & $\sim 10^{32}$ [75]&  \\

J1749.2-2725 & Be? & & 220.38 [8]&
& & 8.5 [2,78] &  $2.6 \cdot 10^{35}$ [2,78] &  \\

J1750-27 &  & & 4.45[8] & 29.8 [8]& & & &   \\

J1820.5-1434 & O9.5-B0Ve [80] & & 152.26 [9]& & &  4.7 [2] &
$9 \cdot 10^{34}$ [2] &  33 [54]\\

1843+00 & B0-B2 & 20.9 [65] & 29.5 [8]& &
& $>10$ [65] & $3\cdot 10^{37}$ [67] & 7 [66]  \\
 &IV-Ve [65]  &  & & & & & &   \\

1845-024 & & & 94.8 [8]& 242.18 [24] & 0.88 [8]& 10 [2] &
$6\cdot 10^{36}$ [2] &  \\

J1858+034 &  & & 221 [8]& & & & &   25 [53]\\

1936+541 & Be [8]& 9.8 [8]& & & & & &   \\

J1946+274 & B0-B1  & 18.6 [64]
& 15.8 [8]& 169.2 [64] & 0.33 [64] & 5 [2] &
$5.4 \cdot 10^{36}$ [2] & 30 [52] \\
 &IV-Ve [64]  &  & & & & & &   \\

J1948+32 & B0e [24]& & 18.76 [24] & 41.7 [24] & $<$0.25 [24] & & &  \\


2030+375 & B0e [24] & 19.7 [8]& 41.8 [24] & 46.03 [8]& 0.41 [24] & 5 [2] &
$ 1 \cdot 10^{38}$ [2] &  36 [70]\\

J2030.5+4751 & B0.5 V-IIIe [8]& & & & & 2.2 [8]& &   \\

J2058+42 & Be? [8]& & 198 [8]& 110 [8]& & 7 [2] &  $2\cdot 10^{36}$ [2] &  36 [70]\\

2103.5+4545 & B0Ve[79] & 14.2 [79] & 358.6 [79] & 12.7 [79] &
$\sim 0.4$ [79] & 6.5 [79] & $ 3 \cdot 10^{36}$ [79] &  \\

2138+568 & B1V-B2Ve [61] & 14.2 [61] & 66.3 [24] & & & 3.8 [2] &  $9.1 \cdot
10^{35}$ [2] &
5-85 [61]  \\
(Cep X-4) & & & & & & & &   \\

2206+543 & B1e [8]& 9.9 [8]& 392 [8]& 9.57 [24] &  &  2.5 [2] &
 $2.5 \cdot 10^{35}$~[2]&  \\

2214+589 & Be [8]& 11 [8]& & & & & &  \\

J2239.3+6116 &  B0 V - B2 IIIe [8]& 15.1 [8] &   1247 [51] &  262.6 [24]  & & 4.4 [8] & $\sim 2.3\cdot 10^{36}$ [51] & 40 [51]     \\

\hline
\end{tabular}
\end{table*}

\subsection{Comments to the tables}

Here we present comments to  the tables.

\noindent
{\bf XTE SMC95.} The source has been revealed during RXTE observations
of the Small Magellanic Cloud, the pulsar was detected in three Proportional
Counter Array (PCA) observations
during an outburst (\citealt{lcp2002}). The source is proposed to be a
Be/neutron star system on the basis
of its pulsations, transient nature and characteristicaly hard X-ray
spectrum. The 2-10 keV X-ray luminosity
implied by observations is $\ga 2\cdot 10^{37}\ \mathrm{erg~s^{-1}}$.
\cite{lcp2002} give the following best fit position: $\alpha=13.36\degr,\,
\delta=-72.821\degr$.

\noindent
{\bf J0032.9-7348.} (RX J0032.9-7348) This source was discovered by
\cite{kp96}
in ROSAT pointed observations made in 1992 December and 1993 April.
\cite{scb99} identified two Be stars within PSPC error circle of  RX
J0032.9-7348.

\noindent
{\bf J0049-732.} (AX J0049-732,  RX J0049.4-7310)
This source was discovered as an X-ray pulsar by \cite{iyk98} with ASCA.
The X-ray flux at 2-10 keV was about $8 \cdot 10^{-13}\ \mathrm{erg~cm^{-2}
s^{-1}}$.
A more likely scenario for  AX J0049-732 is either a Be/X-ray binary or
an anomalous
X-ray pulsar. Direct information to distinguish these two possibilities
can
be obtained by
measuring the pulse period derivative and its orbital modulation.
Two source, No. 427 and No. 430, in the ROSAT PSPC catalogue of
\cite{hfp00}
are possible counterparts of  AX J0049-732. \cite{fph00} searched for
optical
counterparts of these ROSAT sources, and found an emission line object,
possibly a Be star,
at the position of source No. 430, but found no counterpart for source
No.
430. Hence, they
suggest that source No. 427 is more likely to be a counterpart of  AX
J0049-732. However, the
angular separation of these sources of $1.\arcmin 43$ is significantly larger
than
the ASCA error radius.
And \cite{uyi00a} propose that No. 430 is a more likely counterpart.

\noindent
{\bf J0049-729.} (AX J0049-729, AX J0049-728, RX J0049.0-7250, RX
J0049.1-7250,  XTE J0049-729)
This source was discovered with ROSAT (\citealt{kp96}) in pointed
observations.
\cite{yk98a} reported X-ray pulsations in ASCA data of this source. The
X-ray flux in the band
0.7-10 keV was  $1.2 \cdot 10^{-11}\ \mathrm{erg~cm^{-2} s^{-1}}$, with
sinusoidal pulse modulation.
\cite{kp98} suggested the highly variable source RX J0049.1-7250 as
a counterpart.
\cite{scb99} identified two Be stars, one only $3\arcsec$ from the X-ray
position
and one just
outside the error circle given by  \cite{kp96}.
\cite{yit99} reported on the results of two ASCA observations of this
X-ray
source.  The pulse
fraction was $ \sim 70\%$ independent of the X-ray energy.

\noindent
{\bf J0049.4-7323.} (AX J0049.4-7323, AX J0049.5-7323, RX J0049.7-7323)
This X-ray source has been detected 5 times to date, 3 times by the ASCA
observatory (\citealt{yiu00}) and 2 times by the RossiXTE spacecraft.
\cite{uyi00b} reported an ASCA observation which revealed coherent
pulsations of period $755.5 \pm 0.6$ s from
a new source in the Small Magellanic Cloud.
The spectrum was characterized by a flat power-law function
with photon index 0.7 and
X-ray flux $1.1\cdot 10^{-12}\ \mathrm{erg~cm^{-2} s^{-1}}$ (0.7-10 keV).
They noted that the possible Be/X-ray binary RX~J0049.7-7323
(\citealt{hs00}) was located within the ASCA error region.
\cite{ec03} reported on the spectroscopic and photometric analysis of
possible optical counterparts to AX J0049.4-7323. They detected strong
$H_\alpha$ emission from the optical source identified with RX J0049.7-7323
within error circle
for AX J0049.4-7323 and concluded that these are one and the
same object.
They noted that the profile of the curve exhibits a distinct double peak.
This is consistent with Doppler effects which would be expected from a
circumstellar disc viewed in the plane of rotation. There is also definite
V/R asymmetry between the peaks. It is a
compelling evidence for the presence of a Be star.
\cite{cs03} analysed the long term light curve of the optical counterpart
obtained from the MACHO date base. They showed that the optical object
exhibited outbursts every 394~d which they proposed to be the orbital period
of the system. They also showed the presence of a quasi-periodic modulation
with a period $\sim$ 11d which they associated with the rotation of the Be
star disk.
The phase of two RXTE detections is exactly syncronised with the
ephemeris
derived from the optical outbursts. Therefore, as \cite{ce04} concluded,
the period
of 394~d can represent the binary period of a
system with X-ray outbursts
syncronised
with the periastron passage of the neutron star.

\noindent
{\bf 0050-727.} (SMC X-3, H 0050-727, 2S 0050-727, 3A 0049-726, 1H
0054-729,
H 0048-731, 1XRS 00503-727)
SMC X-3 was detected by \cite{ljc77} with SAS 3.
This long-known X-ray  source was not detected by the ROSAT PSPC. But it is
included in the HRI catalogue. \cite{mlt97} reported the detection with
the
RXTE PCA
of an outburst from the X-ray transient SMC X-3 and the discovery of a
period
of $92 \pm 1.5$ s with a complex pulse profile.

\noindent
{\bf J0050.7-7316.} (DZ Tuc, AX J0051-732, RX J0050.6-7315, RX
J0050.7-7316, AX J0051-733, RX J0050.8-7316)
This X-ray source was detected in Einstein IPC, ROSAT PSPC and HRI
archival
data and 18 year
history shows flux variations by at least a factor of 10
(\citealt{iyt99}).
The source was reported as a 323 s pulsar by \cite{yk98b} and
\cite{iyt99}.
Subsequently \cite{c98} identified a 0.7 d optically variable object
within
the ASCA X-ray error circle.
Long term optical data from over 7 years revealed both a 1.4d modulation
and an unusually rapid change in this possible binary period
(\citealt{chl02}).
The system was discussed in the context of
being a normal high mass
X-ray binary by \cite{co00}
who presented some early OGLE data on the object identified by \cite{c98}
and modelled the system
parameters. Coe \& Orosz identified several problems with understanding
this system, primarly that if it was
a binary then its true period would be 1.4 d and it would be an extremely
compact system.
In addition, the combination of the pulse period and such a binary period
violates the Corbet relationship for such systems
(\citealt{c86}). \cite{rl1998} calculated the critical orbital period for
existence of a Be+X-ray pulsar
binary, which is $\sim 10-20$~d. They proposed an explanation for the
lack of Be stars  with accreting
neutron star as companions
with orbital periods less than 10 days as caused by synchronization of
Be star during its evolution.
\cite{chl02} reported on extensive new data sets from both OGLE and
MACHO, as well as on detailed
photometric study of the field. Their results reveal many complex
observational features that are hard to explain
in the traditional Be/X-ray binary model.

\noindent
{\bf J0051-722.} (AX J0051-722,  RX J0051.3-7216)
This source was at first detected as a 91.12 s pulsar in RXTE observations
(\citealt{cml98}) although it
was initially confused with the nearby 46~s pulsar 1WGA J0053.8-7226
(\citealt{bcs98}).
\cite{scb99}  estimated the magnitude of the optical component (Be star) as $V
\sim 15$ from Digitised Sky
Survey images.
The spacing of flares observed from AX J0051-722 suggests an orbital
period of about 120 days (\citealt{isc98}).

\noindent
{\bf 0051.1-7304.} (2E~0051.1-7304, AzV~138)
For this source listed as entry 31 in the Einstein IPC catalogue of
\cite{ww92}
the Be star AzV 138 (\citealt{gh85}) was proposed as an optical counterpart.
2E~0051.1-7304 was not detected in ROSAT observations.

\noindent
{\bf J0051.8-7231.}  (2E 0050.1-7247, RX J0051.8-7231, 1E 0050.1-7247,
1WGA J0051.8-7231)
2E 0050.1-7247 was discovered in Einstein observations. The X-ray
luminosity,
time variability and hard spectrum led \cite{kp96} to suggest a Be/X-ray
binary nature for the source.
\cite{isa95} discovered 8.9 s X-ray pulsations in 2E 0050.1-7247
during a systematic search for periodic signals in a sample of ROSAT PSPC
light curves.
The signal had a nearly sinusoidal shape with a 25-percent pulsed
fraction.
The source was detected several times between 1979 and 1993 at luminosity
levels ranging from
$5 \cdot 10^{34} \mathrm{erg~s^{-1}}$ up to $1.4\cdot 10^{36}
\mathrm{erg~s^{-1}}$ with both
the Einstein IPC and ROSAT PSPC.
The X-ray energy spectrum is consistent with a power-law spectrum that
steepens as the
source luminosity decreases. \cite{isa97} revealed a pronounced
$H_\alpha$ activity from
at least two B stars in the X-ray error circles. These results strongly
suggest that the X-ray
pulsar 2E 0050.1-7247 is in a Be-type massive binary.

\noindent
{\bf J0051.9-7311.} (2E 0050.2-7326, RX J0051.8-7310, AX J0051.6-7311, RX
J0051.9-7311)
This X-ray source was detected by \cite{csm97} during ROSAT HRI
observations
of Einstein IPC source 25 and identified with a Be star by \cite{scc99}.

\noindent
{\bf 0052-723.} (XTE J0052-723)
\cite{cmm01} discovered this transient X-ray pulsar in the direction of
the
Small Magellanic Cloud from RXTE PCA
observations made on 2000 December 27 and 2001 January 5.  Pulsations
were
seen with a period of $4.782 \pm 0.001$ s and with a double-peaked pulse
profile. Spectroscopy of selected optical candidates (\citealt{lcc03})
has
identified
the probable counterpart which is a B0V-B1Ve SMC member exhibiting a
strong,
double peaked $H_\alpha$ emission line.

\noindent
{\bf J0052.1-7319.} (1E 0050.3-7335, 2E 0050.4-7335, RX J0052.1-7319)
The X-ray transient RX J0052.1-7319 was discovered by \cite{lpm99} with
the
analysis
of ROSAT HRI and BATSE data. The object showed a period of 15.3 s
(\citealt{k99a}; \citealt{k99b})
and a flux in the 0.1-2 keV band of  $2.6\cdot 10^{-11}\, \mathrm{erg~
cm^{-2} s^{-1}}$.
\cite{cnc01} reported on the discovery and confirmation of the optical
counterpart of this transient
X-ray pulsar. They found a $V = 14.6$ O9.5IIIe star (a classification as
a
B0Ve star is also possible since the
luminosity class depends on the uncertainty on the adopted reddening).

\noindent
{\bf J0052.9-7158.} (2E 0051.1-7214, RX J0052.9-7158,  XTE J0054-720, AX
J0052.9-7157)
This source was detected as an X-ray transient by \cite{csm97} during ROSAT
HRI observations
of Einstein IPC source 32. The strong variability and the hard X-ray
spectrum imply a Be/X-ray
transient consistent with the suggested Be star counterpart
(\citealt{scc99}).
The X-ray source was detected by ROSAT and is located near the edge of
the
error circle of
XTE J0054-720. The transient pulsar XTE J0054-720 with spin period $\sim 169$
s was discovered with
RXTE (\citealt{lmw98}). \cite{yit03} detected coherent pulsations with
167.8 s period
from AX J0052.9-7157 and determined its position accurately. They found
that AX J0052.9-7157 is
located within the error circle of  XTE J0054-720 and has a variable
Be/X-ray binary,  RX J0052.9-7158, as
a counterpart. From the nearly equal pulse period and the positional
coincidence, they concluded that
the ASCA, ROSAT, and RXTE sources are identical.

\noindent
{\bf J0053.8-7226.} (RX J0053.9-7226, 1WGA J0053.9-7226, 1E 0052.1-7242,
2E 0052.1-7242, RX J0053.8-7226,  1WGA J0053.8-7226,  XTE J0053-724)
This object was  serendipitously discovered as an X-ray source in the SMC
in the ROSAT PSPC archive and also was observed by the Einstein IPC.
Its X-ray properties, namely the hard X-ray spectrum,
flux variability and column density indicate a hard, transient source
with a luminosity of
$3.8\cdot 10^{35}\ \mathrm{erg~s^{-1}}$ (\citealt{bcs01}). XTE and ASCA
observation have confirmed the source to be an X-ray pulsar, with a 46 s
spin period.
Optical observations (\citealt{bcs01}) revealed two possible counterparts to
this
source. Both exhibit strong $H_\alpha$ and weaker $H_\beta$ emission. The
optical colours indicate that both objects are Be stars.
The transient X-ray system XTE J0053-724 was also detected
by RXTE. Pulsations of $46.6 \pm 0.1$ s were observed with a pulse fraction
of about 25\%
(\citealt{l98}). \cite{l98} suggested a possible orbital period of this
Be/X-ray
system of about 139 days which is determined from the periodicity of X-ray
outbursts.

\noindent
{\bf 0053-739.} (SMC X-2, 3A 0042-738, H 0052-739, 2S 0052-739,  H
0053-739, RX J0054.5-7340)
SMC X-2 was one of the first three X-ray sources which were discovered in
the SMC
(\citealt{cdl78}). It was also detected in the HEAO 1 A-2 experiment
(\citealt{mbh79}),
but not in the Einstein IPC survey (\citealt{sm81}). In ROSAT
observations
this transient source was detected only once (\citealt{kp96}). It is
thought to be a Be/X-ray binary, since a Be star was found as its optical
counterpart
(\citealt{mmt79}). In early 2000, the RXTE All-Sky Monitor detected an
outburst
at the position of SMC X-2 (\citealt{cmc01}) and a pulse period of $2.374
\pm 0.007$ s was determined (\citealt{cm00}; \citealt{tky00}). The
source was in low luminosity state during the XMM-Newton observation
(\citealt{sph03}). In
order to estimate the flux upper limit \cite{sph03} used spectral
parameters
derived by \cite{ytk01} from the ASCA spectrum during the outburst.
They obtained an upper limit for the un-absorbed flux of
$1.5\cdot 10^{-14}\, \mathrm{erg~cm^{-2} s^{-1}}$, corresponding
to $L_\mathrm{x} = 6.5\cdot 10^{33}\ \mathrm{erg~s^{-1}}$ ($0.3-10.0$
keV).


\noindent
{\bf 0053+604.} ($\gamma$ Cas, 3A 0053+604, BD+59 144, HD 5394,
LS I $+60\degr133$, 2S 0053+604, 1H 0053+604, 4U 0054+60)
$\gamma$~Cassiopeiae is one of the best known Be stars; it was the first
emission-line star discovered by Angelo Secchi in 1866, and it has spectral
classification of B0 IVe. Its visual magnitude varies between about 3.0 and 1.6, although usually
it stays around 2.5. This object is one of the ROSAT bright sources and also an IRAS source.
$\gamma$ Cas has long been known to be very variable
in  optics  and it is also a moderately strong X-ray source
with a luminosity of the order of $10^{33}\, \mathrm{erg~s^{-1}}$ (\citealt{mws76};
\citealt{wsh82}). Such a luminosity would not be surprising for X-ray emission from an
early type star of spectral type O or B --- some active early type stars have a similar
luminosity (\citealt{cwm94}; \citealt{kmt94}). However, the hardness of the X-ray emission from
$\gamma$ Cas is extraordinarily high in comparison
with usual X-ray emission from early
type stars. If we fit the spectrum with a thermal model the resultant temperature is
roughly 10 keV or more (\citealt{hkh94}; \citealt{mki86}). It is not common for early
type stars, and resembles more the spectra of X-ray pulsars and accreting white dwarf
binaries. There are currently two competing interpretations of the nature of the
observed X-ray emission: one is the accretion of the wind from $\gamma$ Cas onto a
white dwarf companion  and the other one is that it originates from some physical
processes in the outer atmosphere of $\gamma$ Cas itself. Arguments for and against
these two hypotheses are best summarized in studies by \cite{kmi98} and
\cite{rs00}.

\noindent
{\bf J0054.9-7226.} (2E 0053.2-7242, RX J0054.9-7226, 1WGA J0054.9-7226,
SAX J0054.9-7226, RX J0054.9-7227, XTE J0055-724)
RX J0054.9-7226 is known to be an X-ray binary pulsar with a pulse
period of $58.969 \pm 0.001$ s (\citealt{mlc98}; \citealt{scf98}).
\cite{lwc99} have measured the orbital period: 65 d.
In the timing analysis of the XMM-Newton
data, the pulse period was verified to be $59.00 \pm 0.02$~s (\citealt{sph03}). The
optical counterpart, a Be star, is identified with the variable star OGLE J005456.17-722647.6
(\citealt{zsw01}).

\noindent
{\bf J0057.4-7325.} (AX J0057.4-7325)
Six ROSAT observations
have covered the position of AX J0057.4-7325.
Coherent pulsations with a barysentric period of $101.45
\pm 0.07$ s were discovered by \cite{ytk00} with ASCA. The flux
variability, the hard X-ray spectrum, and the long pulse period are
consistent with the hypothesis that AX~J0057.4-7325 is an X-ray binary pulsar
with a companion which is either a Be, an OB supergiant, or a low-mass star.
\cite{ytk00} found only one optical source, MACS J0057-734~10, in the
ASCA error circle. They note that OB supergiant
X-ray binaries in the SMC (only SMC X-1 and EXO 0114.6-7361) are both
located in the eastern wing and this fact may lead us to suspect that
AX J0057.4-7325 would be the third example.

\noindent
{\bf J0058.2-7231.} (RX J0058.2-7231, RX J0058.3-7229)
\cite{scc99} reported the detection of this very weak X-ray source by
ROSAT HRI. Its optical counterpart is a variable Be star in the SMC, OGLE
00581258-7230485 (\citealt{zsw01}).

\noindent
{\bf J0058-720.} (AX J0058-720, RX J0057.8-7202)
The pulse period of  AX J0058-720 was determined from the ASCA data as
$280.4 \pm 0.3$ s (\citealt{yk98b}),
which \cite{sph03} confirmed in the XMM-Newton data: $281.1 \pm 0.2$ s.
The source has been suggested as a Be/X-ray candidate due to the likely
optical counterpart, which is an emission
line object.

\noindent
{\bf J0059.2-7138.}  (RX J0059.2-7138)
The supersoft source RX J0059.2-7138 was detected serendipitously with
the
ROSAT PSPC in 1993 and was seen almost simultaneously be ASCA
(\citealt{hu94}; \citealt{ky96}). Previously, it had failed to be
detected
by either the
Einstein Observatory or EXOSAT in the early 1980s, or in pointed ROSAT
observations
of 1991. The transient nature of this source is clearly established.
The best fit to the X-ray spectrum consists of three components
(\citealt{ky96}): two power laws with indices
0.7 and 2.0 fit the spectrum in the $> 3$ KeV and 0.5-3.0 keV bands
respectively. Furthemore, the
emission is pulsed at levels of $ \sim 35\%$ and $ \sim 20 \%$ in these
respective bands, with a period
of $ \sim 2.7$ s (\citealt{hu94}).
\cite{sc96} identified the probable optical counterpart of this source
with
a 14th-magnitude B1~III emission star
lying within the X-ray error circle.

\noindent
{\bf J0101.0-7206.} (RX J0101.0-7206)
The X-ray transient RX J0101.0-7206  was discovered in the course of
ROSAT
observations of the SMC in
October 1990 (\citealt{kp96}) at a luminosity of $1.3\cdot 10^{36}\
\mathrm{erg~s^{-1}}$.
The source showed a luminosity of  $3\cdot 10^{33}\ \mathrm{erg s^{-1}}$
in the ROSAT
band (0.1-2.4 keV) during two XMM-Newton observations (\citealt{sph03}).
Pulsations with a period of
$304.49  \pm 0.13$ s were discovered in Chandra data
(\citealt{mfl03}). This period could not be verified in the XMM-Newton
observation, because the source was
too faint. \cite{ec03} presented results on the optical analysis of
likely
counterparts, discussing two objects
(Nos. 1 and 4) in the ROSAT PSPC error circle. They conclude that the
optical counterpart is object No. 1 which is
confirmed to be a Be star.

\noindent
{\bf J0101.3-7211.} (RX J0101.3-7211)
The source was detected in ROSAT observations and proposed by \cite{hs00}
as a Be/X-ray candidate.
The optical counterpart (OGLE 01012064-7211187) is a Be star.

\noindent
{\bf J0103-722.} (AX J0103-722,   2E 0101.5-7225,   SAX J0103.2-7209,
CXOU J010314.1-720915, 1E 0101.5-7226)
For the Be/X-ray binary  AX J0103-722 a pulse period of $345.2 \pm 0.1$~s
was determined by \cite{isc98}.
In the XMM-Newton data, pulsations were confirmed with a period of $341.7
\pm 0.4$ s (\citealt{sph03}).
This source was detected with a nearly constant flux  in all the
Einstein, ROSAT and ASCA pointings
which surveyed the relevant region of the SMC.

\noindent
{\bf J0106.2-7205.} (SNR 0104-72.3, RX J0106.2-7205, 2E 0104.5-7221)
SNR 0104-72.3 contains a pointlike X-ray source with a blue optical
counterpart and $H_\alpha$ emission.

\noindent
{\bf 0103-762.} (0107-750)   (1H 0103-762,  H 0107-750)
This source is a very bright UV object with prominent $H_\alpha$ and $H_\beta$
emission.

\noindent
{\bf J0105-722.} (AX J0105-722,  RX J0105.3-7210,   RX J0105.1-7211)
\cite{yk98c}  reported AX J0105-722 as an X-ray pulsar  with a
period of 3.34 s.
From ROSAT PSPC images \cite{fhp00} resolved this source into several
X-ray sources. They combined X-ray,
radio-continuum and optical data to identify the sources: for RX
J0105.1-7211
they proposed an emission line star from the catalogue of
Meyssonier \& Azzopardi in the X-ray error circle as the likely optical
counterpart.
This catalogue contains several known Be/X-ray binaries strongly suggesting
RX J0105.1-7211
as a new Be/X-ray binary in the SMC.

\noindent
{\bf J0111.2-7317.} (XTE J0111.2-7317, XTE J0111-732(?))
The X-ray transient  XTE J0111.2-7317 was discovered by the RXTE X-ray
observatory
in November 1998 (\citealt{clc98a}). Analysis of ASCA observation
(\citealt{cto98b}, \citealt{yit00}) identified this source as a 31 s
X-ray pulsar with a flux in the 0.7-10~keV band of $3.6\cdot 10^{-10}\,
\mathrm{erg~cm^{-2} s^{-1}}$
and $\sim$ 45\% pulsed fraction. The detection was also confirmed from
the BATSE telescope on the CGRO satellite which detected the source in
the hard 20-50 keV band with a flux
ranging from 18 to 30 mCrab (\cite{wf98}).
The source was not detected by ROSAT.
In the X-ray error box of   XTE J0111.2-7317
\cite{cnc01} found
a relatively bright object (V=15.4) which has been classified as a
B0.5-B1Ve star and that
was later confirmed by \cite{chr00} as the most plausible counterpart for
XTE J0111.2-7317. There is also evidence for the presence of a surrounding
nebula, possibly a supernova remnant (\citealt{cnc01}).

\noindent
{\bf 0115+634.} (V635 Cas, 1H 0115+635, 4U 0115+63, 3U
0115+63, 2E 0115.1+6328, H 0115+634, 4U 0115+634)
This source is one of the best studied Be/X-ray systems. This transient
was first reported in the Uhuru satellite survey (\citealt{gmg72};
\citealt{fjc78}),
though a search in the Vela 5B data base revealed that the source had
already
been observed by this satellite since 1969 (\citealt{wrp89}). X-ray
outbursts have been observed from 4U 0115+63
with Uhuru (\citealt{ftj76}), HEAO-1
(\citealt{wdp79}; \citealt{rmh79}), Ginga (\citealt{ttk92}),
CGRO/Batse (\citealt{bcc97}), RXTE
(\citealt{whf99}; \citealt{hc99}; \citealt{crh00}) and reoccur
with a separation times of one to several years. Precise positional
determinations by the SAS 3, Ariel V and HEAO-1
satellites (\citealt{ccl78}; \citealt{jbd78})
were used to identify this system with a heavily reddened
Be star with a visual magnitude $V=15.5$
(\citealt{jkc78}; \citealt{hc81b}). \cite{rcc78} used
SAS 3 timing observations to derive the orbital parameters of this binary
system. Due to the fast rotation of the neutron star
centrifugal inhibition of accretion prevents the onset of X-ray emission
unless the ram pressure of accreted material reaches
a relatively high value. Magnetic field of the neutron star
is $1.3\cdot 10^{12}\ \mathrm{G}$ (\citealt{mmn99}).
Pulse fraction was obtained in a model-dependent way in the range 20-50 keV
(see Harmon et al. 2004\nocite{hwf04} for details and references).

\noindent
{\bf J0117.6-7330.} (RX J0117.6-7330)
This X-ray transient was discovered by the PSPC on
board ROSAT (\citealt{crw96}; \citealt{crw97}).
\cite{s99} conducted spectroscopic and photometric observations of the
optical companion of the
X-ray transient RX J0117.6-7330 during a quiescent state. The primary
component was identified as a B0.5~IIIe star.
\cite{mfh99} reported on the detection of pulsed, broadband, X-ray
emission
from this transient source.
The pulse period of 22 s was detected by the ROSAT/PSPC instrument and by
the Compton Gamma-Ray
Observatory/BATSE instrument. The total directly measured X-ray
luminosity
during the ROSAT observation
was $1.0\cdot 10^{38}\ \mathrm{erg~s^{-1}}$. The pulse frequency increased
rapidly during the outburst with a peak spin-up
rate of $1.2\cdot 10^{-10}\ \mathrm{Hz~s^{-1}}$ and a total frequency
change of 1.8\%. The pulsed percentage
was 11.3\% from 0.1-2.5 keV, increasing to at least 78\% in the 20-70 keV
band.
These results established RX J0117.6-7330 as a transient Be binary
system.

\noindent
{\bf J0146.9+6121.} (V831 Cas, 2S 0142+61, RX J0146.9+6121,  LS I
$+61\degr 235$)
RX J0146.9+6121 is an accreting neutron star with a 25 min spin period,
the
longest known
period of any X-ray pulsar in a Be-star system.
This fact was realized (\citealt{msn93}) only after the re-discovery of
this source in the ROSAT
All Sky Survey and its identification with the 11th magnitude Be star
LS I $+61\degr 235$ (\citealt{mbb91}).
Indeed the 25 min periodicity had already been discovered with EXOSAT
(\citealt{wmg87}), but it
was attributed to a nearby source 4U 0142+614.
The optical star is probably a member of the open cluster NGC 663 at a
distance of
about 2.5 kpc (\citealt{tce91}). For this distance, the 1-20 keV
luminosity
during the EXOSAT
detection in 1984 was $\sim 10^{36}\, \mathrm{erg~s^{-1}}$
(\citealt{msn93}).
All the observations of RX J0146.9+6121 carried out after its
re-discovery
yielded lower luminosities,
of the order of a few $10^{34}\ \mathrm{erg~s^{-1}}$ , until an
observation
with the Rossi XTE satellite
showed that in July 1997 the flux started to rise again (\cite{ham98}),
though not up to the level of the
first EXOSAT observation.

\noindent
{\bf 0236+610.} (V615 Cas, 2E 0236.6+6101, LS I $+61\degr303$, 1E
0236.6+6100, RX J0240.4+6112)
LS I $+61\degr 303$ is a radio emitting X-ray binary which exhibits radio
outbursts every 26.5 d. The radio outburst peak and the outburst phase
are
known to vary over
a time scale of $\sim$ 4 yr (\citealt{gxb89}; \citealt{g99}).  The 26.5~d
period is believed to
be the orbital period. \cite{hc81} confirmed the radio period by analysis
of
three-year observation of radial velocity. They concluded that the
optical
spectrum corresponds
to a rapidly rotating B0 V star. The 4 yr modulation
has been discovered on the basis of continued radio monitoring.

\noindent
{\bf 0331+530.} (BQ Cam, EXO 0331+530, V 0332+53)
EXOSAT observed three outbursts from V0332+53 between 1983
November and 1984 January, leading to the discovery of the 4.4 s spin
period and a sudden decrease of luminosity at the end of $\sim 1$ month
long recurrent outbursts. The latter result was interpreted as the
onset of the centrifugal barrier (\citealt{swd85}; \citealt{swr86}). An
upper
limit of $\sim 5 \cdot 10^{33}\, \mathrm{erg~s^{-1}}$  to the source
quiescent
emission (1--15 keV) was derived on that occasion with the EXOSAT Medium
Energy
Detector. Doppler shifts in pulse arrivals indicate that the
pulsar is in orbit around a Be star with a period of 34.3 days and
eccentricity 0.3 (\citealt{swd85}). Observations during a
subsequent outburst with Ginga led to the discovery of a cyclotron
line feature corresponding to  $3\cdot 10^{12}\ \mathrm{G}$ magnetic
field
(\citealt{mkk84}).
BeppoSAX and Chandra observations  allowed to study this transient
at the faintest flux levels thus far (\citealt{csi02}).
\cite{csi02} concluded that the quiescent emission of this X-ray
transient
likely originates
from accretion onto the magnetospheric boundary of the neutron star in
the
propeller
regime and/or from deep crustal heating resulting from pycnonuclear
reactions during the outbursts.


\noindent
{\bf 0352+309.} (X~Per, HD~24534, 3A~0352+309, 2E~0352.2+3054,
H~0352+309,
4U~0352+30,
4U~0352+309, 1H~0352+308, 2A~0352+309, H~0353+30, HD~24534, 3U~0352+30)
The X-ray source 4U~0352+309 is a persistent low luminosity pulsar in a
binary system
with the Be star X~Persei (X~Per). Its 837 s pulsation period was
discovered with the
UHURU satellite (\citealt{wms76}; \citealt{wms77}), and is still one of
the
longest periods of any
known accreting pulsar (\citealt{bcc97}, and references therein). X Per
is
a bright and highly
variable star with a visual magnitude that ranges from $\sim 6.1$ to
$\sim
6.8$
(\citealt{mbf74}; \citealt{rlt97}). The spectral class has been estimated
to be O9.5~III to B0~V
(\citealt{s82}; \citealt{frc92}; \citealt{lrr97}).
Based on spectroscopic parallax, distance estimates range from $700 \pm
300$ pc up to $1.3 \pm 0.4$ kpc
(\citealt{frc92}; \citealt{lrr97}; \citealt{rlt97}; \citealt{twr98}).
The X-ray luminosity varies on long timescales (years) from
$\sim 3\cdot 10^{35}\ \mathrm{erg~s^{-1}}$ to
$\sim 5\cdot 10^{34}\ \mathrm{erg~s^{-1}}$ (for assumed distance of 1.3
kpc; \citealt{rcf93}).
\cite{dlp01} have determined a complete orbital ephemeris of the system
using data from the Rossi
X-ray Timing Explorer (RXTE).
\cite{chg01} have discovered a cyclotron resonant scattering feature at
29
keV in the
X-ray spectrum of 4U~0352+309 using observation taken with the RXTE.
The cyclotron resonant scattering feature energy implies a magnetic
field strength at the polar cap of $3.3\cdot 10^{12}\ \mathrm{G}$.

\noindent
{\bf J0440.9+4431.} (RX J0440.9+4431,  VES 826)
RX J0440.9+4431/BSD 24-491 was confirmed as an accreting
Be/X-ray system following the discovery of X-ray pulsations, with
barycentric pulse
period of $202.5 \pm 0.5$ s from RXTE observations (\citealt{rr99}).

\noindent
{\bf J0501.6-7034.} (RX J0501.6-7034,  2E 0501.8-7038, 1E 0501.8-7036, HV
2289, CAL 9)
This Einstein and ROSAT variable source was identified with a Be star by
\cite{scf94}.
Later \cite{sch96} identified this star with HV~2289, a known variable
with
a large
amplitude of variability.

\noindent
{\bf J0502.9-6626.} (RX J0502.9-6626, CAL E)
The X-ray source  RX J0502.9-6626 was originally detected by the Einstein
observatory
(\citealt{cch84}) at a flux of $\sim 3\cdot 10^{36}\, \mathrm{erg~
s^{-1}}$. The source
was detected three times with the ROSAT PSPC at luminosities
$\sim 10^{35} - 10^{36}\, \mathrm{erg~s^{-1}}$ and once with the HRI
during
a bright
outburst at $4\cdot 10^{37}\, \mathrm{erg~s^{-1}}$ (\cite{scm95}). During
the outburst, pulsations at
4.0635 s were detected. The identification of this source with the Be
star
[W63b]~564 = EQ~050246.6-663032.4
(\citealt{w63}) was confirmed by \cite{scf94}.

\noindent
{\bf J0516.0-6916.} (RX J0516.0-6916)
The identification of this source with a Be-star is unclear.
In several observations the source  did not display any characteristics
of
Be  behaviour, however,
Schmidtke et al. (1999) classify it as a Be-star.

\noindent
{\bf J0520.5-693.} (RX J0520.5-6932)
This X-ray source has been observed at a low X-ray luminosity
($5\cdot 10^{34}$ erg s$^{-1}$)
in early 90-s by ROSAT (\citealt{scf94}).
The light curve of the optical counterpart exhibits significant
modulation with a period of 24.5 d,
which is interpreted as the orbital period (\citealt{cnb01}).
A spectral type of O9V was proposed for the optical counterpart.
In a recent paper \cite{ecg2004} present new optical and IR data
and archive BATSE data on the ouburst.

\noindent
{\bf J0529.8-6556.} (RX J0529.8-6556, RX J0529.7-6556)
The transient X-ray source RX J0529.8-6556 was detected during one single
outburst as a 69.5-s X-ray pulsar by \cite{hdp97}, who identified it with
a relatively bright blue star showing weak $H_\alpha$ emission.

\noindent
{\bf 053109-6609.2.} (EXO 053109-6609.2, RX J0531.2-6609,  RX
J0531.2-6607,
EXO 0531.1-6609)
This source was discovered by EXOSAT during deep observations og the LMC
X-4
region in 1983 (\citealt{pbp85}). It was detected again in 1985 by the
SL2
XRT experiment.
The lack of detection in EXOSAT observations made between these dates
demonstrates the
transient nature of the source. The companion is optically identified
with
a Be star (\citealt{hdp95}).
\cite{bsr98} reported a timing analysis of the Be transient X-ray binary
EXO 053109-6609.2 in outburst
observed with BeppoSAX. The pulsed fraction is about constant in the
whole
energy range.
The source shows pulsations from 0.1 up to 60 keV.
In the MECS (Medium Energy Concentrator Spectrometer) pulse profile in
the
1.8-10.5 keV band the pulsed fraction
is $0.54 \pm 0.05$. In the LECS (Low Energy Concentrator Spectrometer)
pulse profile (the 0.1-1.8 keV band), the
main pulse is still evident, while the interpulse is more broadened, and
pulsed fraction is
$0.78 \pm 0.28$. The PDS (Phoswich Detection System) pulse profile (15-60
keV energy band)
still shows a double-peaked structure (pulsed fraction is $0.64 \pm
0.16$)
in phase with the previous ones.
Although the statistics is poor, the pulsed fraction does not seem to
decrease with energy (\citealt{bsr98}).

\noindent
{\bf J0531.5-6518.} (RX J0531.5-6518)
This source was detected with the ROSAT PSPC in June 1990
(\citealt{hp99}).
The source is probably variable, since other pointings failed to detect
it.
The optical counterpart is
probably a Be star coming back from an extended disk-less phase
(\citealt{nc02}).

\noindent
{\bf J0535.0-6700.} (RX J0535.0-6700)
This source was observed by the ROSAT PSPC at a luminosity
$\sim 3\cdot 10^{35} \mathrm{erg~s^{-1}}$ (\citealt{hp99}). Its
positional coincidence with an optically variable star in the LMC (RGC28
in
\citealt{rgc88}) is very good.
RGC28 is an early-type Be star and likely it is the optical counterpart to
RX~J0535.0-6700 (\citealt{nc02}). The star displays periodic variability
in its I-band lightcurve at 241 d, which \cite{rgc88} originally believed
to be
the period of a Mira variable. \cite{hp99} suggested that this variability
can
be related to the orbital period.

\noindent
{\bf 0535-668.} (RX J0535.6-6651, 1A 0538-66, 1A 0535-66)
This source was discovered by the Ariel 5 satellite in June 1977, during
outburst in which the flux peaked at $\sim 9\cdot 10^{38} \mathrm{erg~
s^{-1}}$
(\citealt{wc78}). When active, 1A 0535-66 displays very bright short
X-ray outbursts separated by 16.6 days, which is believed to be the
orbital
period. The optical
counterpart experiences drastic changes in the spectrum, with the
appearance of strong
P-Cygni-like emission lines, and brightening by more than 2 mag in the
$V$
band (\citealt{cbd83}).
The Be star has a $V$ magnitude of $\sim 14.8$ during the X-ray quiescent
periods. The magnitude
reaches a peak of 12.5 mag during the X-ray outbursts.
Detection of a 69-ms pulsation in the X-ray signal has been reported only
once (\citealt{sbe82}).
Further X-ray observations of outbursts were made by \cite{sss80} using
the
HEAO 1
satellite. The X-ray outbursts were found to last up to at least 14
days or to be as short as a few hours. 1A~0535-66 in its largest
outbursts (\citealt{sss80}) has luminosity around $10^{39}\, \mathrm{erg
s^{-1}}$.
ROSAT (\citealt{mh93}) and ASCA observations (\citealt{csc95}) have
revealed low-level
outbursts with luminosities of $4\cdot 10^{37}\, \mathrm{erg ~s^{-1}}$
and
$2\cdot 10^{37}\, \mathrm{erg ~s^{-1}}$ in the two ROSAT observations
and
$\sim 5.5\cdot 10^{36}\, \mathrm{erg~s^{-1}}$ in the ASCA observation.
Due to the low count rate and sampling frequency it was not possible to
determine whether the 69 ms pulsations were present in the data.
The ratio of $L_\mathrm{max}$ to $L_\mathrm{min}$ in soft X-rays is $ >1000$.

\noindent
{\bf 0535+262.} (V725 Tau, HD 245770, 1A 0535+26, 1H 0536+263, 3A
0535+262,
BD+26 883, 4U 0538+26, 1A 0535+262, H 0535+262)
The transient A 0535+26 is one of the best studied Be/X-ray binaries.
This source was discovered in 1975 by Ariel 5 (\citealt{res75}) and
showed
a periodicity at 104 s indicating the presence of a highly magnetized
neutron star.
The optical counterpart was later identified with the Be star HDE~245770
(\citealt{lcj79})
allowing the classification of the source as a Be/X-ray binary.
The pulsed fraction is 20\% at 30-40 keV and increases significantly with
energy,
reaching 100\% at 100 keV (\citealt{ffm85}).
Magnetic field of the neutron star is $4.3\cdot 10^{12}\ \mathrm{G}$
(\citealt{mmn99}).

\noindent
{\bf 0544-665.} (H 0544-665, H 0544-66)
This source was discovered with the HEAO-1 scanning modulation collimator
by \cite{jbd79}.
The brightest object within the X-ray error circle was found to be a
variable B0-1 V star
(\citealt{kte83}) but no emission lines have been observed in its
spectrum
to identify
it as a Be star. \cite{kte83} published photometry which showed a
negative
correlation
between optical magnitudes and colour indices, typical of Be stars whose
variability is
due to variations in the circumstellar disc. \cite{scb99} suggested that
the object may be
a Be star in the state of low activity.

\noindent
{\bf J0544.1-710.} (RX J0544.1-7100, AX J0544.1-7100, AX J0548-704,
1WGA J0544.1-7100, 1SAX J0544.1-7100)
This source is a transient X-ray pulsar ($P=96$ s) with the hardest
X-ray spectrum observed
by ROSAT in the LMC (\citealt{hp99}). The observations of the optical
counterpart were
presented by \cite{cnb01}, who found it to display large variability in
the
$I$-band lightcurve
and $H_\alpha$ in emission. An approximate spectral type of B0~Ve was
proposed.

\noindent
{\bf 0556+286.} (4U 0548+29, 1H 0556+286) The X-ray source was detected by
HEAO1, earlier probably observed by UHURU 4U 0548+29 (\citealt{wood1984}).
No detection was made after that.
A Be-star is known in this direction.

\noindent
{\bf J0635+0533.} (SAX J0635+0533)
Discovered by BeppoSAX (\citealt{kph1999}).
\cite{z2002} gives the spectral classification of the optical counterpart
as B0.5 IIIe.
X-ray luminosty is
$(9-35) \cdot 10^{35}$~erg~s$^{-1}$ (2-10 keV) for $d=2.5-5$~kpc
(\citealt{kph1999}).
Bolometric luminosity (0.1-40 keV) was estimated to be
$1.2 \cdot 10^{35}$~erg~s$^{-1}$ for $d=5$~kpc (\citealt{cmn2000}).
Pulse fraction was obtained by BeppoSAX (2-10 keV).
The source can be identified with the gamma-ray source 2EG J0635+0521.
Low luminosity together with very fast rotation propose that
the NS has a low magnetic field (see discussion in Cusumano et al.
2000\nocite{cmn2000}).

\noindent
{\bf 0726-260.} (4U 0728-25, 3A 0726-260, V441 Pup, 1H 0726-259, LS 437)
Detected by many experiments (UHURU, HEAO1, Ariel 5, ROSAT, RXTE).
Pulse fraction was estimated
as $(I_\mathrm{max}-I_\mathrm{min})/(I_\mathrm{max}+I_\mathrm{min})$
from the graph in \cite{cp1997} (RXTE 2-20 keV).
The spectral and photometrical analysis of this source led \cite{nrb1996} to conclude
that the primary is an O8-9Ve star.

\noindent
{\bf 0739-529.} (1H 0739-529 ) Detected by HEAO1 (\citealt{wood1984}).

\noindent
{\bf 0749-600.} (1H 0749-600) Detected by HEAO1 (\citealt{wood1984}).
Situated in the open cluster NGC 2516 (\citealt{liu2000}).

\noindent
{\bf J0812.4-3114.} (RX J0812.4-3114, V572 Pup, LS 992)
RX~J0812.4-3114 was discovered by \cite{mhd1997} during a search
for high-mass X-ray binaries by cross-correlating SIMBAD OB star catalogs
with low Galactic latitude sources from the ROSAT all-sky survey. This X-ray
source thus has an identified optical counterpart, the Be star LS~992, and so
it was suspected that this source belongs to the Be/X-ray binaries.
\cite{rnb2001} classify it as B0.2 IVe.
The X-ray light curve of LS 992/RX~J0812.4-3114
is characterised by 31.88 second
pulsations, while the X-ray spectrum is best
represented by an absorbed power-law
component with a exponentially cut-off (\citealt{rr1999a}).
In December 1997 the source made a transition from 
a quiescent state to a flaring state (\citealt{cp2000}),
in which regular flares separated by 80 day intervals 
were detected with the All-Sky
Monitor (ASM) onboard the Rossi X-ray Timing Explorer. \cite{cp2000} attributed
the origin of these flares to the periastron passage of the neutron star, 
hence this periodicity was naturally associated with the orbital period.
\cite{cp2000} have found strong evidence for the presence 
of a $\sim 80$ day period in the ASM light curve of RX~J0812.4-3114. 
By comparison with other Be star X-ray binaries,
the time of maximum flux is likely to coincide with periastron passage of a
neutron star. The orbital period of $\sim 80$ days combined with the $\sim 32$
second pulse period is consistent with 
the correlation between orbital and pulse
period that is found for the majority of 
Be/neutron star binaries (\citealt{c86}).

\noindent
{\bf 0834-430.} (GS 0834-430)
The hard X-ray transient GS~0834-430 was discovered by the WATCH
experiment on board GRANAT in 1990 at a flux level of about 1 Crab in
the 5-15 keV energy band (see \citealt{whs1997}). The source was later
observed by GINGA (\citealt{mak90a}; \citealt{mak90b}) 
and ROSAT as a part of the All
Sky Survey (\citealt{hpb90}). 
The pulsations at a period of 12.3 s were
observed during the GINGA, ROSAT and ART-P observations (\citealt{mak90c};
\citealt{ade1992}; \citealt{hpb90}; \citealt{gs91}).
GS 0834-43 was also monitored by BATSE) 
between April 1991 and July 1998. In particular
seven outbursts were observed from April 1991 
and June 1993 with a peak and intra-outburst
flux of about 300 mCrab and $ < 10$ mCrab, 
respectively (\citealt{wfh1997}). The recurrence
time of 105-115 days was interpreted 
as the orbital period of the system. However, no further
outbursts have been observed since July 1993 
either with CGRO/BATSE and the All Sky
Monitor  on board  RXTE. All these findings suggested
that GS 0834-43 is a new Be-star/X-ray binary system with an
eccentric orbit (\citealt{wfh1997}).
Based on both photometric and 
spectroscopic findings \cite{icc2000} concluded that optical
counterpart of this X-ray pulsar is most 
likely a B0-2 V-IIIe star at a distance of 3-5 kpc.
Pulse fraction was obtained by BATSE (20-50 keV).

\noindent
{\bf J1008-57.} (GRO J1008-57 ) Discovered by BATSE in 1993.
Pulse fraction $\sim$~60\% was
obtained by ROSAT (0.1-2.4 keV) (\citealt{hwf04}).
High-energy data (BATSE: 20-70 keV) gives nearly the same value about 67\%
(see Harmon et al. 2004\nocite{hwf04}).
Orbital period is uncertain. An estimate of 247.5 days comes from
the best fit of BATSE data (\citealt{no2001}). Other (earlier) estimates
were about 135 days (\citealt{liu2000}).

\noindent
{\bf 1036-565.} (3A 1036-565, 1A 1034-56, )
Probably the same object as J1037.5-5647.

\noindent	  
{\bf J1037.5-5647.} (LS 1698, RX J1037.5-5647) 
Discovered by ROSAT in 1997.
Probably the same source as 4U1036-56/3A1036-565.
The source was observed in quiescence
(\citealt{rr99}). $L_\mathrm{min}=1.1\cdot 10^{34}$~erg~s$^{-1}$.
Pulse fraction was obtained by RXTE (3-20 keV).

\noindent
{\bf 1118-615.} (1A 1118-615, 1A 1118-616, WRAY 15-793, 2E 1118.7-6138) 
The hard X-ray transient A 1118-615 was discovered serendipitously in 1974 
by the Ariel-5 satellite (\citealt{esw75}) during an observation 
of Cen~X-3 (4U~1119-603). The same series of observations revealed 
pulsations with a period of $405.3 \pm 0.6$ s (\citealt{isb1975}). However, in 
the initial announcement of the discovery of the pulsations, they were 
wrongly attributed to an orbital period, suggesting that A~1118-615 consisted 
of two compact objects (\citealt{isb1975}).
This hard X-ray transient underwent a major outburst only twice: 
in 1974, when it was discovered by Ariel-5 satellite, and from December 1991 
to February 1992 (\citealt{bcc97}).
The source was observed by \cite{mjp88} using the Einstein and EXOSAT
observatories in 1979 and 1985 respectively. On both occasions a weak 
signal was detected confirming that low-level accretion was occurring.
The correct optical counterpart was identified as the Be star 
He~3-640/Wray~793 by \cite{ci75}. 
The primary has been classified as O9.5IV-Ve (\citealt{jic81}), 
with strong Balmer emission lines 
indicating the presence of an extended envelope. 
According to \cite{vgp92}, the exact classification is complicated by 
many faint absorption and emission lines (mostly of Fe~II), but the overall 
spectrum is found to be similar to that of the optical counterparts to other 
known Be/X-ray sources.
The source was observed by \cite{cp85} at UV wavelengths using the IUE 
satellite. They confirmed the identification of the counterpart and reported
prominent UV lines characteristic of a Be star.
Despite the large observational efforts made during last years and mainly 
after the 1991-1992 outburst, the Hen3-640/1A~118-615 system is still poorly 
understood. The orbital period of the system is unknown. Corbet's pulse
period/orbital period diagram (\citealt{c86}) gives an orbital period
estimate of $\sim 350$ days.

\noindent	  
{\bf 1145-619.} (V801 Cen, 2S 1145-61, 2S 1145-619, 2S 1145-62, LS 2502, 3U
1145-61,
4U 1145-62, 4U 1145-619, 4U 1145-61, 3A 1145-619, 2E 1145.5-6155, H
1147-62, H 1145-619) 
Initially observed by UHURU (together with
1145.1-9141). Two sources were distinguished by Einstein observatory
(HEAO2).
In \cite{liu2000} the optical counterpart was classified as B1 Vne.
Pulse fraction was obtained by BATSE (20-50 keV).


\noindent	  
{\bf 1249-637.} (1H 1249-637, 2E 1239.8-6246, BZ Cru) 
Detected by HEAO1 (\citealt{wood1984}).
Probably a WD accretor.

\noindent
{\bf 1253-761.} (1H 1253-761) Detected by HEAO1 (\citealt{wood1984}).
Probably a WD accretor.

\noindent
{\bf 1255-567.} (1H 1255-567, $\mu^2$ Cru) 
Detected by HEAO1 (\citealt{wood1984}).

\noindent
{\bf 1258-613.} (GX 304-1, 4U 1258-61, V850 Cen, H 1258-613, 2S 1258-613,
3A 1258-613 ) 
Discovered by UHURU.
In \cite{z2002} classified as B0.7Ve.


\noindent
{\bf 1417-624.}  (2S 1417-624, 2S 1417-62, 4U 1416-62, 2E 1417.4-6228, 3A
1417-624, H 1417-624)
The X-ray source 2S 1417-62 was detected by SAS-3 in 1978 (\citealt{ank1980}).
Analysis of the SAS 3 observations showed evidence of $\sim$
57 mHz pulsations (\citealt{kad81}). 
Einstein and optical observations identified a Be star companion at
a distance of 1.4-11.1 kpc (\citealt{gpc84}). 
From the timing analysis of BATSE 
observations orbital parameters were determined
and a correlation was found between spin-up rate and
pulsed flux (\citealt{finger1996}). 
Orbital period and eccentricity of the source were found to be 
42.12 days and 0.446 respectively.

\noindent
{\bf J1452.8-5949.} (1SAX J1452.8-5949) 
1SAX J1452.8-5949 was discovered during a BeppoSAX 
galactic plane survey in 1999 (\citealt{oop1999}).
Coherent pulsations were detected with a barycentric period of 
a $437.4 \pm 1.4$ s. The X-ray properties and lack of an obvious optical 
counterpart are consistent with a Be star companion at a distance of 
between approximately 6 and 12 kpc.
Pulse fraction is high. It was determined in the BeppoSAX band 1.8-10 keV.
Be/X-ray systems display a correlation between their spin and orbital 
periods (\citealt{c86}) which in this case implies an
orbital period of $>$200 days for 1SAX~J1452.8-5949.

\noindent
{\bf J1543-568.} (XTE J1543-568) 
The transient X-ray source XTE J1543-568 was discovered 
by RXTE in 2000 (\citealt{icm2001}).
A subsequent pointed PCA observation revealed a pulsar with a period 
of $27.12 \pm 0.02$~s. Later 
the pulsar was found in earlier data from 
BATSE on board the Compton Gamma-Ray Observatory.
The orbital period is $75.56 \pm 0.25$ d. The mass function and position
in the pulse period versus orbital period diagram are consistent with
XTE J1543-568 being a Be/X-ray binary. The eccentricity is less than 0.03,
so it is among the lowest for twelve Be/X-ray binaries
whose orbits have now been well measured.
This confirms the suspicion that small kick velocities of neutron
stars in HMXBs are more common than expected (\citealt{icm2001}).
Pulse fraction (RXTE) slightly depends on energy (from 2 to 20 keV).

\noindent
{\bf 1553-542.} (2S 1553-542, 2S 1553-54, H 1553-542)
The X-ray source 2S 1553-542 was discovered during observations with
SAS 3 in 1975 (\citealt{kra1983}).
Pulse fraction was determined by SAS-3 (2-11 keV).

\noindent
{\bf 1555-552.} (1H 1555-552, LS 3417, RX J155422.2-551945, 2E 1550.3-5510,
1E 1550.4-5510)
Detected by HEAO1 (\citealt{wood1984}).

\noindent
{\bf J170006-4157.}  (AX J170006-4157, AX J1700-419, AX J1700.1-4157)
This source was discovered and observed three times between 1994 and
1997 by ASCA (\citealt{torii1999}). Significant 
pulsations with P = $714.5 \pm 0.3$ s 
were discovered from the third observation.
The X-ray spectrum is 
described by a flat power-law function with a photon index of —0.7. Although
the spectrum could also be fitted by thermal models, the temperature obtained 
was unphysically high. The hard spectrum suggests that the source is a neutron
star binary pulsar similar to X Persei (4U~0352+309), but it cannot be 
completely excluded the possibility that it is a white dwarf binary.
Not marked as a Be-candidate in \cite{liu2000}.
Pulse fraction in the range 0.7-10 keV was
determined from the graph in \cite{torii1999}.

\noindent
{\bf J1739-302.} (XTE J1739-302, AX J1739.1-3020) 
This source was discovered in an observation of the black hole 
candidate 1E 1740.7-2942 with the proportional counter array (PCA) 
of the Rossi X-Ray Timing Explorer (\citealt{smith1998}).
Luminosity estimated for a 2-100 keV range with an assumption, that the
source is at the Galactic center.
\cite{smith1998} tentatively identified XTE J1739-302 as a Be/NS binary 
because its spectral shape is similar to that of these systems: 
a gradual steepening over the 2-25 keV range.

\noindent
{\bf J1739.4-2942.} (RX J1739.4-2942)
Discovered by ROSAT (\citealt{motch1998}).
Probably identical with GRS 1736-297.

\noindent
{\bf J1744.7-2713.} (RX J1744.7-2713, HD 161103, V3892 Sgr, LS 4356)
Discovered by ROSAT (\citealt{mhd1997}).
Luminosity was estimated for an energy range 0.1-2.4 keV.
Pulsed fraction was taken from paper by \cite{hwf04}. It has been
obtained by BATSE in the range 20-40 keV.

\noindent
{\bf J1749.2-2725.}  (AX J1749.2-2725)
Discovered by ASCA (\citealt{torii1998}).
Not marked as a Be-candidate in \cite{liu2000}.

\noindent
{\bf J1750-27.} (GRO J1750-27, AX J1749.1-2639)
GRO J1750-27 is the third of the new transient accretion-powered
pulsars discovered using BATSE.
A single outburst from GRO J1750-27 was observed with BATSE 
(see \citealt{scott1997}).
Pulsations with a 4.45~s period were discovered on 1995 July 29 
from the Galactic center region as part of the BATSE all-sky pulsar monitoring 
program (\citealt{bcc97}).
An orbit with a period of 29.8 days was found by \cite{scott1997}. 
Large spin-up rate,
spin period and orbital period together 
suggest that accretion is occurring from a disk 
and that the outburst is a ``giant'' one typical 
for a Be/X-ray transient system. 

\noindent
{\bf 1820.5-1434.} (AX 1820.5-1434)
This X-ray source was discovered in 
1997 by  ASCA  (\citealt{kth1998}).
Pulsations with a period $\sim$ 152 s were 
detected in the 2-10 keV flux of the source with
a pulsed fraction of $\sim$ 50\%. The pulse fraction is not energy dependent.
Both timing and spectral properties of AX 1820.5-1434 are typical for
an accretion-driven X-ray pulsar.
\cite{icp2000} proposed O9.5-B0Ve 
star as an optical counterpart of the pulsar.

\noindent
{\bf 1843+00.} (GS 1843+00)
The transient X-ray source GS 1843+00 was discovered during 
the Galactic plane scan near the Scutum region by X-ray detectors
on board the Ginga satellite (\citealt{ttp89}). Coherent pulsations 
with a period of about 29.5 s were observed with a very small peak-to-peak 
amplitude of only 4 per cent of the average flux.
Pulse fraction was obtained by BATSE (20-50 keV).
Luminosity estimates are the follwoing: 
1) $2\cdot 10^{36}$~erg~s$^{-1}$ (20-200~keV, 10
kpc) (\citealt{m1999}); 2) $3\cdot 10^{37}$~erg~s$^{-1}$
(0.3-100 keV, 10 kpc) (\citealt{pss2000}).

\noindent
{\bf 1845-024.} (2S 1845-024, GS 1843-02, 4U 1850-03, 1A 1845-02, 1H
1845-024, 3A 1845-024, GRO J1849-03)
The pulsar GS 1843-02 was discovered by 
Ginga in 1988 (\citealt{makino88}) during a galactic
plane scan conducted as part of a search for transient pulsars
(see \citealt{finger1999}). 
The same source is known as GRO J1849-03. 
X-ray outbursts occur regularly every 242 days.
\cite{finger1999} presented 
a pulse timing analysis that shows that the 2S 1845-024 outbursts 
occur near the periastron passage. 
The orbit is highly eccentric
(e = $0.88 \pm 0.01$) with a period of 
$242.18 \pm 0.01$ days. 
The orbit and 
transient outburst 
pattern strongly suggest that the pulsar
is in a binary system with a Be star. From 
the measured spin-up rates and 
inferred luminosities \cite{finger1999} concluded that 
an accretion disc is 
present during outbursts.

\noindent
{\bf J1858+034.} (XTE J1858+034)
The hard X-ray transient XTE J1858+034 was discovered with the
RXTE All Sky Monitor  in 1998 (\citealt{rl98}). The spectrum was 
found to be hard similar 
to spectra of X-ray pulsars. Observations were made 
immediately after this with 
the Proportional Counter Array (PCA) of the RXTE and 
regular pulsations with 
a period of $221.0 \pm 0.5$ s were discovered (\citealt{tcm1998}).
The pulse profile is found to be nearly sinusoidal with a pulse fraction of 
$\sim$ 25\%. From 
the transient nature of this source 
and pulsations they suggested that this is a Be-X-ray 
binary. The position of the X-ray source 
was refined by scanning the sky around the source
with the PCA (\citealt{mlc98}).
From the RXTE target of opportunity (TOO) 
public archival data of the observations 
of XTE J1858+034, made in 1998, \cite{pr1998} have discovered the presence 
of low frequency QPOs.
Pulse fraction was obtained by RXTE (2-10 keV).

\noindent
{\bf 1936+541.} (1H 1936+541) 
Detected by HEAO1 (\citealt{wood1984}).

\noindent
{\bf J1946+274.} (XTE J1946+274, GRO J1944+26, 3A 1942+274, SAX
J1945.6+2721)
Pulse fraction obtained by
Indian X-ray Astronomy Experiment - IXAE (2-18 keV).
\cite{chr2002} present a data on cyclotron feature in the spectrum
of J1946+274 which corresponds to the field $\sim 3.9\cdot 10^{12}$~G.
\cite{wfc2003} propose a distance $9.5\pm 2.9$~kpc basing on a
correlation between measured spin-up rate and flux.

\noindent
{\bf J1948+32.} (GRO J1948+32, GRO J2014+34, KS 1947+300)
Discovered by BATSE (see Chakrabarty et al. 1995\nocite{chak1995}).
Galloway et al. (2004)\nocite{gml2004} presented results which can
indicate
a glitch in that system.


\noindent
{\bf 2030+375.} (EXO 2030+375, V2246 Cyg)
EXO 2030+375 was discovered in 1985 May with EXOSAT satellite during 
a large outburst phase (\citealt{pws89}). This outburst was first 
detected at a 1-20 keV energy band and its luminosity is close to the 
Eddington limit (assuming 5 kpc distance to the source) for 
a neutron star (\citealt{psf85}).
The X-ray emission of the transient pulsar EXO 2030+375 is modulated
by $\sim$ 42 s pulsations and periodic $\sim$ 
46 days Type I outbursts, that are 
produced at each periastron passage of the neutron star, i.e. when 
the pulsar interacts with the disk of the Be star. 
Not marked as a Be-candidate in \cite{liu2000}.
See a detailed description in \cite{wfc2002}.
Pulse fraction was obtained by BATSE in the range 30-70 keV 
(see \citealt{hwf04}).

\noindent
{\bf J2030.5+4751.} (RX J2030.5+4751, SAO 49725) Discovered by ROSAT
(see \citealt{mhd1997}). 
This object is marked as a likely Be/X-ray candidate
in \cite{liu2000}, but not in many other papers.
The pointing data show that the X-ray source is relatively hard. 
The $L_\mathrm{x} /L_\mathrm{bol}$  
ratio is close to $3\cdot 10^{-6}$. This is rather
strong evidence for an accreting compact object around
SAO 49725 (\citealt{mhd1997}).

\noindent
{\bf J2058+42.} (GRO J2058+42)
GRO J2058+42, a transient 198~s X-ray pulsar. It was discovered by BATSE 
during a ``giant'' outburst
in 1995 (see \citealt{wfh1998}). The pulse period decreased from 198 to 196~s 
during the 46 day outburst.
BATSE observed five weak outbursts from GRO J2058+42 that were spaced by
about 110 days.
The RXTE All-Sky Monitor detected eight weak outbursts with approximately
equal durations and intensities.
GRO J2058+42 is most likely a Be/X-ray binary that appears to produce
outbursts at periastron and apastron. No optical counterpart has been
identified to date (see however \citealt{ct96}), and no X-ray source was present in
the error circle in archival ROSAT observations (\citealt{wfh1998}).
Pulse fraction was obtained by BATSE in the range 20-70 keV
(see \citealt{hwf04} for details).

\noindent
{\bf J2103.5+4545.} (SAX J2103.5+4545)
SAX J2103.5+4545 is a transient HMXB pulsar
with a $\sim 358$ s pulse period discovered
with the WFC on-board BeppoSAX during an outburst in 1997 (\citealt{hzh98}).
Its orbital period of 12.68 days
has been found with the RXTE during the 1999
outburst (\citealt{b2000}). The likely optical counterpart, a Be star with a
magnitude V=14.2, has been recently discovered (\citealt{rm2003}).
During the outburst in 1999 \cite{b2002} for the first time observed  with
RXTE a transition from the spin-up phase to the spin-down regime,
while the X-ray flux was declining.
\cite{i2004} observed a soft
spectral component (blackbody with a temperature of
1.9 keV) and a transient 22.7 s
QPO during a XMM-Newton observation performed in 2003.

\noindent
{\bf 2138+568.} (GS 2138-56, Cep X-4, V490 Cep, 1H 2138+579, 4U 2135+57, 3A
2129+571)
The X-ray source Cep X-4 was discovered with a transient high level X-ray
flux in 1972 by OSO-7
(\citealt{ubw73}). The source was not detected again
till 1998 when a new outburst was detected by GINGA.
During these observations coherent 66~s
pulsations were discovered revealing
an X-ray pulsar with a complex X-ray spectrum including a possible 30~keV
cyclotron absorption
feature (\citealt{kkt91}, \citealt{mmk1991}).
Cep X-4 has been associated with a Be star that lies within the ROSAT error
box.
A cyclotron line was detected by \cite{mmk1991},
it corresponds to the magnetic field $B=2.3\cdot 10^{12} (1+z)$~G.
Pulse fraction strongly depends on energy and is highly variable with
time from nearly 0 up to $>80$\%
(see \citealt{wfs1999b}).
RXTE pulse fraction is decreasing with intensity.

\noindent
{\bf 2206+543.}  (3U 2208+54, 4U~2206+54, 1H 2205+538, 1A 2204+54, 3A
2206+543)
The hard X-ray source 4U 2206+54 was first detected by the Uhuru satellite
(\citealt{gmg72}).
The source
is included in the Ariel V catalogue as 3A 2206+543 (\citealt{wmf81}).
4U 2206+54 has been detected by all satellites that have pointed at it and
has never been observed to undergo an outburst.
\cite{sfg84} used the refined position from the HEAO-1 Scanning Modulation
Collimator to identify the optical counterpart with the early-type star BD
+53$\degr$ 2790.
From their photometry, they estimated that the counterpart was a B0~-~2e main
sequence star,
and therefore concluded that the system was a Be/X-ray binary.
\cite{crp00} have announced the detection of a $9.570 \pm 0.004$ d
periodicity in the X-ray lightcurve.
If this is the binary period, then it would be the shortest known for
a Be/X-ray binary --- unless the $\sim$ 1.4 d periodicity
in the optical lightcurve of RX J0050.7-7316 (\citealt{co00}).
Optical and ultraviolet spectroscopy of the optical component
BD +53$\degr$ 2790
show it to be a very peculiar
object, displaying emission in H I, He I and He II lines and variability in
the intensity of many lines of metals (\citealt{nr01}). 
Strong wind troughs in the UV resonance lines suggest a large mass loss
rate. These properties might indicate
that the star displays at the same time the Of and Oe phenomena or even
a hint of the possibility that it could be a 
spectroscopic binary consisting of two massive stars in addition to the
compact object (\citealt{nr01}). With  
all certainty there is
an O9.5V star in the system which is probably a mild Of
star, and which likely feeds the compact
object with its stellar wind (\citealt{nr01}).
See also recent data and discussion in (\citealt{cp2001}).
These authors confirm the orbital period of $\sim 9.6$ days.
This value is surprisingly short if one takes into account
long spin period of the NS (see fig.~2, where this system is definitely
displaced from the normal trend). Spin period was not detected
in many observations. Corbet \& Peele (2001)\nocite{cp2001}
discuss several possibilities
other than Be/X-ray interpretation including an accreting WD. 
Nearly perfect alignment between magnetic and spin axis is also a
possibility.

\noindent
{\bf 2214+589.} (1H 2214+589)
Detected by HEAO1 (Wood et al. 1984\nocite{wood1984}).
This object is mentioned in \cite{liu2000} as a
Be-candidate. However, it is not mentioned in many lists of Be/X-ray
systems
(for example in \citealt{z2002}). Not much is known about this source.

\noindent
{\bf J2239.3+6116.} (3A 2237+608, SAX J2239.3+6116, SAX J2239.2+6116, 3U
2233+59, 4U 2238+60)
Discovered by BeppoSAX (see \citealt{isc2001}).
SAX J2239.3+6116 is an X-ray transient which often recurs with 
a periodicity of 262 d (\citealt{ihe2000}). 
Because of the Be-star nature of the likely optical counterpart the
periodicity may be 
identified with the orbital period of the binary. 
Pulse fraction was determined from the graph in
\cite{isc2001}
as $(I_\mathrm{max}-I_\mathrm{min})/(I_\mathrm{max}+I_\mathrm{min})$. 
It corresponds to the energy range $\sim 1$~--~10~keV.
$L_\mathrm{max}$ corresponds to the distance 4.4~kpc and
the highest flux $10^{-9}$~erg~cm$^{-2}$s$^{-1}$
in the energy range 2-28~keV (\citealt{isc2001}).


\section{Graphs and discussion}

In this section we give a graphical representation of the data.
In the first figure we show a usual period -- luminosity dependence.
If luminosity is proportional to $\dot M$~--~an accretion rate~--~ then
for each value of $L$ it is possible to determine a critical period,
$P_\mathrm{A}$ (see details on the magnetorotational evolution of neutron stars
for example in \citealt{l92}). 
It is determined by an equality of the magnitospheric radius
to the corotation radius, so $P_\mathrm{A}$ depends also on the magnetic field
of a neutron star. If a spin period of a NS is shorter than $P_\mathrm{A}$
then accretion rate is significantly reduced, and the NS is at the stage of
{\it propeller}. Lines for $P_\mathrm{A}$ for two values of the magnetic field
are shown in the figure. Situation can be more complicated for low accretion 
rates when the so-called {\it subsonic propeller} stage becomes important.
In that case for a neutron star it is necessary  to slow down to a new
critical period $P_\mathrm{crit}$. Lines for this quantity are also shown
(see figure caption for other details).

In the second figure we present the so-called "Corbet diagram" (\citealt{c86}).
For most of Be/X-ray binaries correlation between spin and orbital periods is
strong, so that this dependence is even used to estimate orbital periods
when only spins are known.

Pulse profiles contain a lot of important information about accretors.
The simplest characteristic of a pulse profile is its pulse fraction.
Therefore it is useful to look if there are correlations of this quantity with
other parameters.
In the next three figures we plot pulse fraction
vs. spin period and luminosity.
Pulse fraction usually is energy dependent.
In the tables for several objects we give maximal and minimal values
of measured pulse fraction. In the figures 3 and 4 in these cases
we plot two points corresponding to upper and lower limits given in the table.

As it is clear form the fig.~3
pulse fraction is not correlated with the maximal luminosity.
Also there is no correlation between spin period and pulse fraction at least
for not very long ($<$300~s) spins.
If there is any correlation between spin period and pulse fraction
for long  periods is less
clear due to small statistics (see figs. 4 and 5).
It would be interesting to have larger statistics
on pulse fraction for systems with long orbital periods in connection with
discussion in  the paper by \cite{lh96}.

The observational number distribution of Be/X-ray binaries over orbital
characteristics is shown in figs. 6 and 7. It is seen that Be-systems do
not have orbital periods longer than one year.
There is a lack of systems with periods
10--20 days. As it was shown in the
paper by \cite{rl1998} the lack of short-period
Be/X-ray binaries can be explained by the effect of tidal synchronization in
binaries. The peak of the observed number distribution of Be/X-ray
systems over eccentricities falls in the range $0.4-0.5$.
In order to get a better agreement with the observed
parameters of Be/X-ray binaries there is no necessity of high kicks.
Moderate recoil velocities of the order  50 km s$^{-1}$ are enough
(see \citealt{rl1998}).

\begin{figure}
\vbox{\psfig{figure=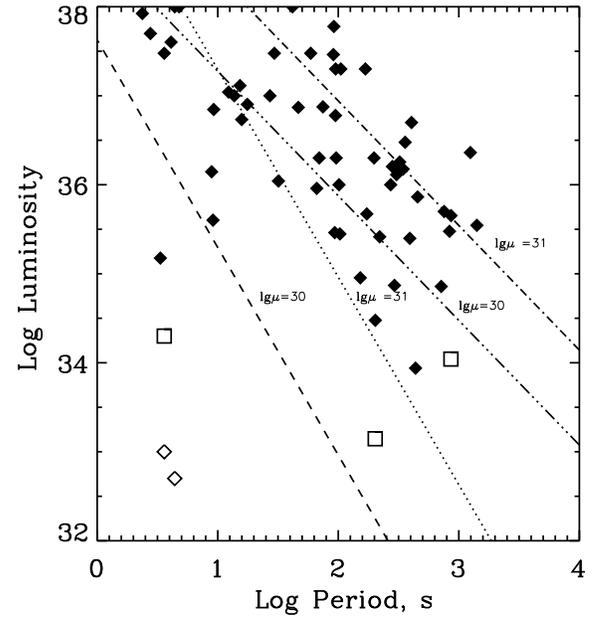,width=9.0cm}}
\caption[]{Period~--~Luminosity.
Open symbols correspond to the quiescent state
of the X-ray pulsar.
Squares represent three sources in quiescence from
which pulsations were observed (4U 0115+63 -- Campana et al. 2001;
RX J0440.9+4431 and RX J1037.5-564 -- Reig, Roche 1999b).
Open diamonds show objects without pulsations in quiescence,
which are supposed to be
in the {\it propeller} state (4U 0115+63 and V0332+53 -- Campana et al.
2002).
The graph is artifitially cutted at log~$p=0$ and
log~$L=38$.
So, here we do not plot
two systems with the most fastly rotating NSs: J0635+0533 (small luminosity)
and 0535-668 (large luminosity).
Dashed and dotted lines correspond to the
critical period,
$P_\mathrm{A}=2^{5/14}\pi (GM)^{-5/7} (\mu^2/\dot M)^{3/7}$, for two values
of the magnetic moment, $\mu=10^{30}$~G~cm$^3$
and $10^{31}$~G~cm$^3$. The
two dashed-dotted lines correspond to subsonic propeller~--~accretor
transition for the same two values of the magnetic moment which
occurs at
$P_\mathrm{crit}=81.5 \mu_{30}^{16/21} L_{36}^{-5/7}$ according to Ikhsanov
(2003).
We note that the multiplicative coefficient in Ikhsanov's formula is
larger than in the classical formula of
Davies, Pringle (1981) by a factor $\sim7.5$.
}
\label{fig:pl}
\end{figure}

\begin{figure}
\vbox{\psfig{figure=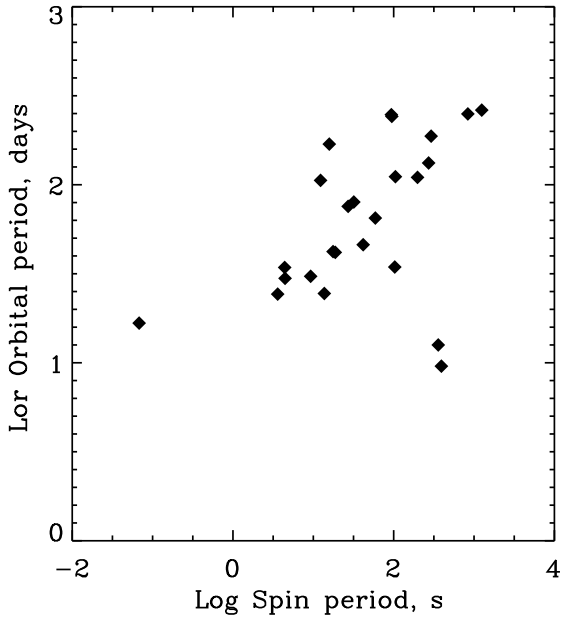,width=9.0cm}}
\caption[]{Spin period~--~Orbital period.  Three displaced systems are:
2206+543 (large spin and short orbital periods),
2103.5+4545 (large spin and short orbital periods) and 0535-668 (very short
spin period).
}
\label{fig:pporb}
\end{figure}

\begin{figure}
\vbox{\psfig{figure=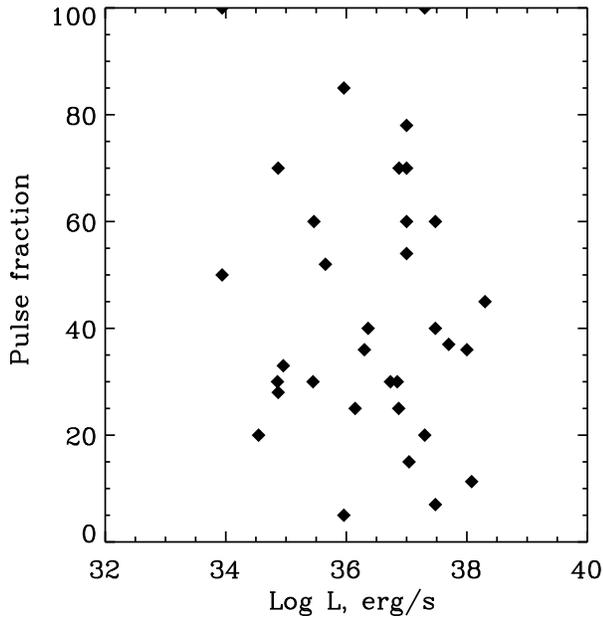,width=9.0cm}}
\caption[]{Luminosity~--~Pulse fraction.
}
\label{fig:pfl}
\end{figure}

\begin{figure}
\vbox{\psfig{figure=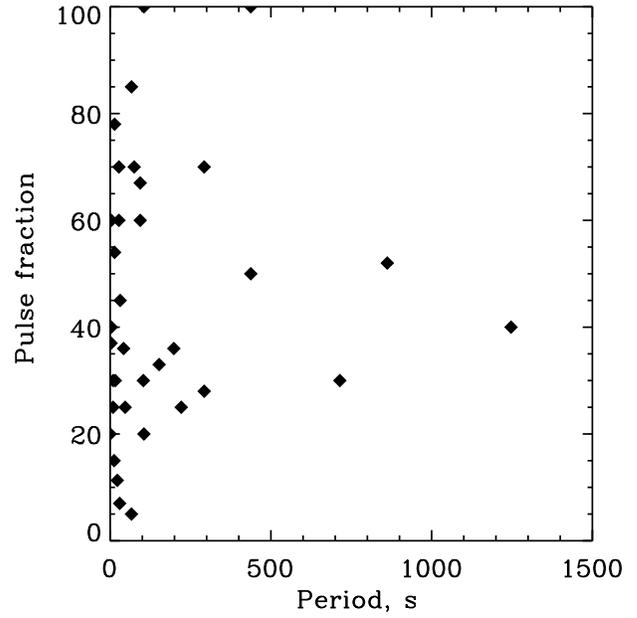,width=9.0cm}}
\caption[]{Spin period~--~Pulse fraction. If there is deficit of high and low
pulse fraction at long spin periods is unclear due to small statistics.
}
\label{fig:ppf}
\end{figure}

\begin{figure}
\vbox{\psfig{figure=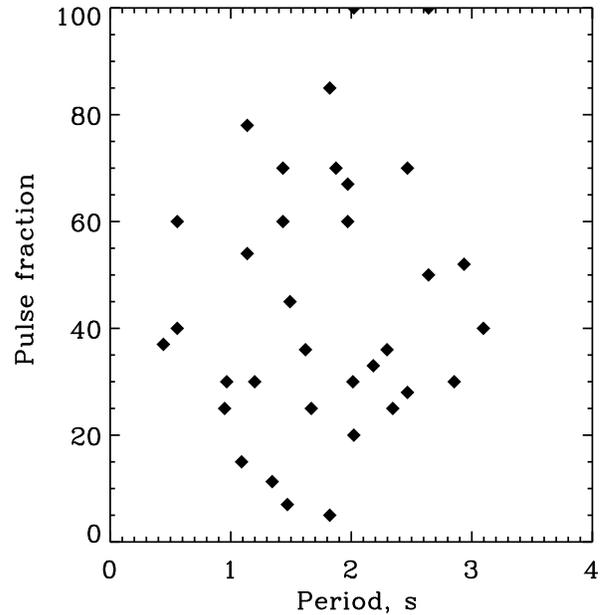,width=9.0cm}}
\caption[]{Spin period~--~Pulse fraction. Data is the same as in the
previous figure, but spin periods are given in a logarithmic scale.
}
\label{fig:lppf}
\end{figure}

\begin{figure}
\includegraphics[width=9.0cm]{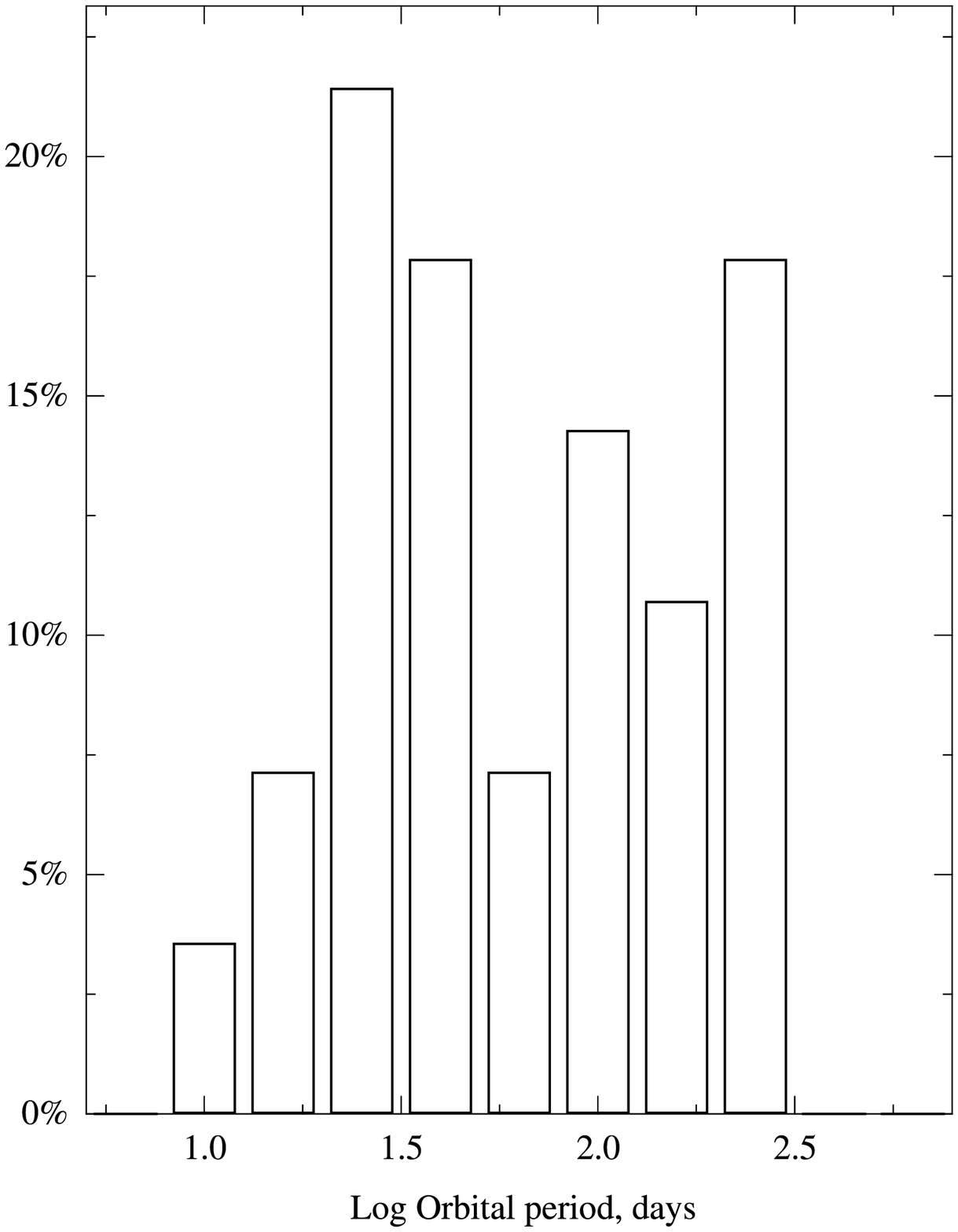}
\caption[]{The observational number distribution of Be/X-ray binaries over
orbital period.}
\label{fig:gist_porb}
\end{figure}

\begin{figure}
\includegraphics[width=9.0cm]{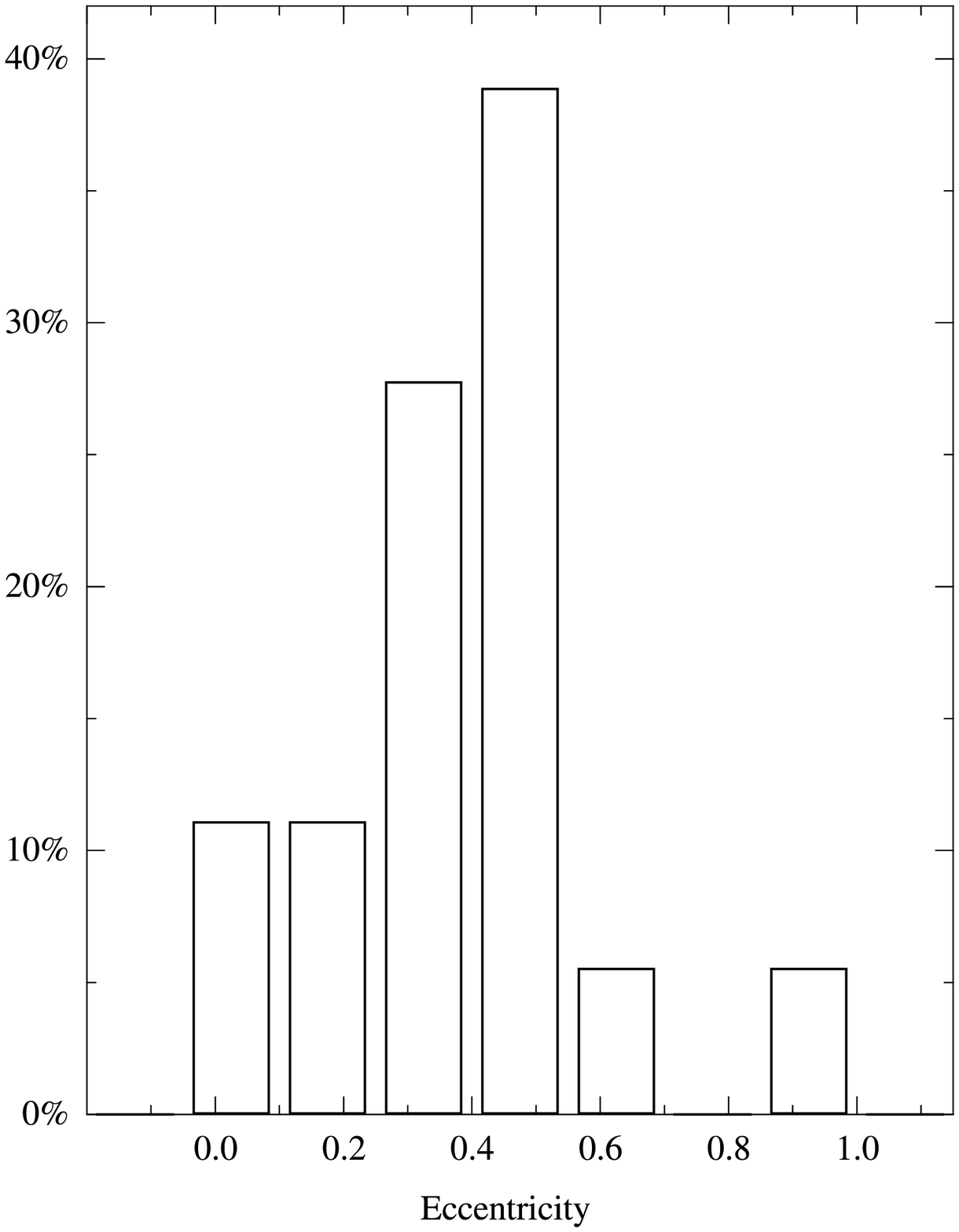}
\caption[]{The observational number distribution of Be/X-ray binaries over
orbital eccentricity}
\label{fig:gist_ecc}
\end{figure}



\begin{acknowledgements}

 The work was supported by the
Russian Foundation for Basic Research (RFBR)
grant 03-02-16068.

\end{acknowledgements}

\end{document}